\documentclass[aps,twocolumn,amssymb,showpacs]{revtex4-1}
\usepackage{graphicx}
\usepackage{multirow}
\usepackage[breaklinks,colorlinks,urlcolor=blue,citecolor=blue]{hyperref}
\usepackage{mathrsfs}
\usepackage{amsmath}
\usepackage{soul}
\usepackage{color}
\usepackage[normalem]{ulem}

\begin{document}

\title{Hot and Dense Homogeneous Nucleonic Matter Constrained by Observations,
  Experiment, and Theory}
\author{Xingfu Du$^{1}$}
\author{Andrew W. Steiner$^{1,2}$}
\author{Jeremy W. Holt$^{3}$}
\affiliation{$^{1}$Department of Physics and Astronomy, University of
  Tennessee, Knoxville, TN 37996, USA}
\affiliation{$^{2}$Physics Division, Oak Ridge National Laboratory, Oak
  Ridge, TN 37831, USA}
\affiliation{$^{3}$Cyclotron Institute and Department of Physics and
  Astronomy, Texas A\&M University, College Station, Texas 77843, USA}

\begin{abstract}
  We construct a new class of phenomenological equations of state for
  homogeneous matter for use in simulations of hot and dense matter in
  local thermodynamic equilibrium. We construct a functional form
  which respects experimental, observational and theoretical
  constraints on the nature of matter in various density and
  temperature regimes. Our equation of state matches (i) the virial
  coefficients expected from nucleon-nucleon scattering phase shifts,
  (ii) experimental measurements of nuclear masses and charge radii,
  (ii) observations of neutron star radii, (iii) theory results on the
  equation of state of neutron matter near the saturation density, and
  (iv) theory results on the evolution of the EOS at finite
  temperatures near the saturation density. Our analytical model
  allows one to compute the variation in the thermodynamic quantities
  based on the uncertainties in the nature of the
  nucleon-nucleon interaction. Finally, we perform a correction to
  ensure the equation of state is causal at all densities,
  temperatures, and electron fractions. 
\end{abstract}


\maketitle

\section{Introduction}

The equation of state (EOS) of dense nucleonic matter is a central
microphysical input required for numerical simulations of
core-collapse supernovae~\cite{Hix14ei} and neutron star
mergers~\cite{Baiotti17bn}. These simulations probe baryon densities
($n_B$) up to a few nucleons per fm$^3$, temperatures ($T$) up to 100
MeV, and a wide range of electron fractions ($Y_e$). This
three-dimensional space is often described through an EOS table which
relates the free energy or pressure of the system to the thermodynamic
variables describing the ambient environment. These tables, in turn,
are built from modeling the interactions between nucleons, nuclei, and
potentially exotic particles that may appear at high densities.

Over this large three-dimensional $(n_B,Y_e,T)$ space, there are
several different physical regimes each constrained by different
observables and theoretical approaches. The first regime, zero
temperature nuclear matter at nuclear saturation density, is closely
connected to nuclear masses, charge radii, giant resonances, and other
laboratory observables. Global fits to experimental data have been
performed with Skyrme (e.g., Ref.~\cite{Kortelainen14}) and covariant
mean-field models (e.g., Ref.~\cite{RocaMaza14ca}). The second regime,
cold neutron matter below nuclear saturation density, is difficult to
probe experimentally but is well-constrained by theoretical
calculations based on semi-phenomenological nuclear forces or
microscopic chiral effective field theory-based
interactions~\cite{Gandolfi15nm,Hebeler15nf,Coraggio13}. The third
regime, strongly-interacting high-temperature matter, is best
described by interactions and many-body approaches similar to those
applied to cold neutron matter near saturation
density~\cite{Mukherjee06vt,Wellenhofer15to,Carbone13sg,Drischler15an}.
The fourth regime, low-density and high-temperature matter that is
nearly non-degenerate, is best described by the virial expansion. The
equation of state in this regime is determined from nucleon-nucleon
scattering phase shifts~\cite{Horowitz06,Horowitz06b}. Finally,
neutron-rich matter at densities above twice saturation density is
most strongly constrained by observations of neutron star masses and
radii, particularly the observation of neutron stars with $M\simeq
2M_{\odot}$~\cite{Demorest10,Antoniadis13}.

There are several currently available tabulated equations of state.
The first set of EOSs, developed by Lattimer and Swesty
(LS)~\cite{Lattimer91}, was constructed in the single-nucleus
approximation and based on three different non-relativistic Skyrme
interactions. Two of the three Skyrme interactions have nuclear
incompressibilities ($K$) far outside of modern constraints
\cite{Youngblood99,Shlomo06}. The third, with $K=220$ MeV, has a
combination of symmetry energy ($S$) and slope of the symmetry energy
($L$) that are only slightly outside of current constraints
(\cite{Fischer14,Lattimer14co,Holt17,Tews17}). The LS EOS with $K=220$
MeV also produces a 2 $\mathrm{M}_{\odot}$ neutron star and is still
important for simulations of core-collapse supernovae and neutron star
mergers. The second set of EOSs came from H. Shen et al.~\cite{Shen98}
(also using the single-nucleus approximation) and was based on the NL3
relativistic mean-field Lagrangian. The values of $K$ and $L$ for NL3
are much larger than current neutron star obervations~(see the
analysis in e.g., \cite{Steiner10te}) and nuclear
theory~\cite{Hebeler10} suggest.

While the single-nucleus approximation is sufficient to describe the
bulk thermodynamics, it does not in general accurately describe the
composition~\cite{Burrows84,Hix03,Botvina05,OConnor07,Arcones08,Souza09b}
and the associated weak reaction rates. More modern EOS tables often
include a more complete nuclear distribution as a result. The third
set from G.\ Shen et al.~\cite{Shen10} includes an EOS table based on
a more modern relativistic mean-field model,
``FSUGold''~\cite{ToddRutel05}, and goes beyond the single nucleus
approximation to include a full distribution of nuclei in nuclear
statistical equilibrium (NSE). This model has values of $K$, $L$ and
$S$ that are within recent constraints from experiment and (in later
versions) produces a neutron star maximum mass larger than 2
$\mathrm{M}_{\odot}$. These EOSs based on FSUGold include more modern
nuclear physics input, including a proper treatment of nearly
non-degenerate matter that matches the virial expansion. The fourth
set of EOS tables was based on the work of Hempel and
others~\cite{Hempel12} that built upon several nucleon-nucleon
interactions which produce reasonable values of $K$, $S$ and $L$ and
generate 2 $\mathrm{M}_{\odot}$ neutron stars, including FSUGold,
DD2~\cite{Typel10}, IUFSU~\cite{Fattoyev10}, SFHo~\cite{Steiner13cs}
and SFHx~\cite{Steiner13cs}. The latter two interactions were designed
to simultaneously match laboratory nuclei and give neutron star radii
that match astronomical observations of neutron
stars~\cite{Steiner10te}. More recently, several EOSs have been added
to the CompOSE (CompStar Online Supernovae Equations of State)
database~\cite{Typel13cc}, including an EOS with
hyperons~\cite{Banik14}.

In this work, we construct a phenomenological free energy density that
is consistent with observational and theoretical constraints in the
five aforementioned physical regimes. This is in contrast to works
which attempt to describe matter over the entire density and
temperature range with a single detailed model of the nucleon-nucleon
interaction. Many previous works proceed this way using a Skyrme-based
or relativistic mean-field model to describe matter at all densities
and temperatures. The principal problem is that these models are
guaranteed to work well only for isospin-symmetric nuclear matter at
zero temperature. Extrapolating these models to other density and
temperature regimes may lead to inaccurate EOS results or may
introduce unphysical correlations between the nature of matter across
different regimes. For example, given a Skyrme model it is common to
observe that the nuclear incompressibility is correlated with the
maximum mass of neutron stars. Such a correlation has little physical
meaning, however, since the neutron star maximum mass is determined by
interactions in high-density matter that likely have little similarity
to nucleons in the laboratory (see a similar argument in
Ref.~\cite{Prakash88}). We avoid extrapolations where possible, but
some extrapolation will still be required where experimental and
theoretical guidance is lacking.

Our second advance is in the treatment of uncertainties. The most
relevant parameters which describe the uncertainties in different
density and temperature regimes are not clearly related. The virial
expansion provides a clear path forward for describing uncertainties
at low-density and high-temperature, but higher-order virial terms are
not necessarily useful for quantifying uncertainties at higher
densities. In this work, through the construction of a
phenomenological model one can vary uncertainties in different regimes
independently, without spoiling agreement elsewhere.

\section{Method}

The EOS table is constructed by combining an EOS for homogenous
nucleonic matter consisting only of neutrons and protons. The EOS is
written in the form of the Helmholtz free energy (including only the
contribution from nucleons) $f_{\mathrm{np}}(n_B,Y_e,T)$. In the
discussion below, we remove the nucleon rest mass contributions from
the free energy densities and chemical potentials and use a tilde when
these rest mass contributions are included, i.e.
\begin{eqnarray}
  \tilde{f}_{\mathrm{np}}(n_B,Y_e,T) &\equiv& f_{\mathrm{np}}(n_B,Y_e,T)
  \nonumber \\ && +
  n_B \left[ \left(1-Y_e\right) m_n + Y_e m_p
    \right],
\end{eqnarray}
where $m_n$ and $m_p$ are the neutron and proton masses. When
electrons are included, their rest mass contribution to the free
energy is also included. We ignore muons and exotic charged particles
at higher densities. Thus the proton fraction, $x_p$, and electron
fraction, $Y_e$ are always equal.

\subsection{Virial expansion and homogeneous nucleonic matter}

The virial expansion is a model-independent way of computing the
pressure of matter at low densities and high
temperatures~\cite{Horowitz06,Horowitz06b,Shen11}. It is an expansion in
powers of the fugacity, $z_i$, of particle $i$ defined by
\begin{equation}
  z_i = \exp \left( {\mu}_{i,\mathrm{vir}} / T \right) ,
\end{equation}
where ${\mu}_{i,\mathrm{vir}}$ denotes the nucleon chemical
potential. In matter consisting only of
neutrons and protons, the first-order terms in the pressure,
proportional to $z_n$ and $z_p$, consist of the classical
non-interacting contribution to the pressure. The coefficients of the
second-order terms (second-order virial coefficients) in the virial
expansion can be obtained directly from nucleon-nucleon scattering
phase shifts. Third-order virial coefficients are not well known.
Nuclear statistical equilibrium implies that the fugacity of nuclei
can be written in terms of the neutron and proton fugacities
\begin{equation}
  z_{(Z,N)} \propto z_n^{N} z_p^{Z} \, .
\end{equation}
Thus when the neutron and proton fugacities are nearly equal the
contribution of deuterons comes at second order in the virial
expansion and the contribution from alpha particles comes at fourth
order. Two-body scattering between nucleons and alpha particles
contributes at fifth order in the virial expansion. In this work,
because the third-order virial coefficients are not well-known,
third-order and higher terms are ignored. 

Second-order terms in the virial expansion affect the description
of homogeneous nucleonic matter. In order to ensure that
the free energy matches the virial result at low densities
and high temperatures, the free energy density is written as
\begin{eqnarray}
  f_{\mathrm{np}}(n_B,x_p,T) &=& f_{\mathrm{virial}}(n_n,x_p,T) g \nonumber
  \\ && + f_{\mathrm{deg}}(n_B,x_p,T) (1-g) \, ,
  \label{eq:fnp}
\end{eqnarray}
where $f_{\mathrm{virial}}$ is the virial free energy density,
$f_{\mathrm{deg}}$ is the free energy density when either
the neutrons or protons are sufficiently degenerate so
that the virial expansion is a poor approximation. The
function $g$ is defined by
\begin{equation}
  g \equiv 1/(1+3z_n^2 + 3z_p^2) \, .
  \label{eq:g}
\end{equation}
This definition ensures that $(1-g)f_{\mathrm{deg}}$ appears as a
third- or higher-order correction to the free energy density in the
virial expansion as long as $f_{\mathrm{deg}}$ is at least linear in
the fugacity at low densities (we verify this below). The value of $g$
is 1 only when $z_n$ and $z_p$ are both sufficiently small. The
numerical coefficient 3 was chosen to ensure a positive entropy in
the entire region in $(n_B,Y_e,T)$ space for which the pressure of our
EOS is positive. The relationship between the fugacities and the
densities is
\begin{eqnarray}
  n_n &=& 2 \lambda^{-3}\left[ z_n + 2 z_n^2 b_n(T)+ 2 z_n z_p
    b_{pn}(T)\right] \nonumber \\
  n_p &=& 2 \lambda^{-3}\left[ z_p + 2 z_p^2 b_n(T) + 2 z_n z_p
    b_{pn}(T)\right] \, .
  \label{eq:virden}
\end{eqnarray}
These equations are solved for the fugacities in order to compute the
free energy density from the virial expansion. The quantity $\lambda
\equiv \left[4 \pi/(m_n T+m_p T)\right]^{1/2}$ is the average nucleon
thermal wavelength, the quantity $b_n(T)$ is the second neutron virial
coefficient, and the quantity $b_{pn}(T)$ is the virial coefficient
describing the interaction between neutrons and protons.

The virial coefficients $b_n(T)$ and $b_{pn}(T)$ are determined by
scattering phase shifts. Analytical fits can be employed, similar to
those in Ref.~\cite{Horowitz12}, but previous fits for $b_n$ employ
functional forms which diverge for $T\rightarrow 0$. We perform an
alternate fit, constraining the zero temperature behavior to match
that expected from a finite-range expansion, which will be correct
when the density is sufficiently small. The values of the virial
coefficients at high temperature are not well-known, so we arbitrarily
constrain the fits so that the virial coefficients give the value
expected for noninteracting fermions at $T=150$ MeV. While very hot
and nearly nondegenerate matter is present in simulations, it is
unlikely to strongly affect the dynamics.

\begin{figure}
  \includegraphics[width=3.1in]{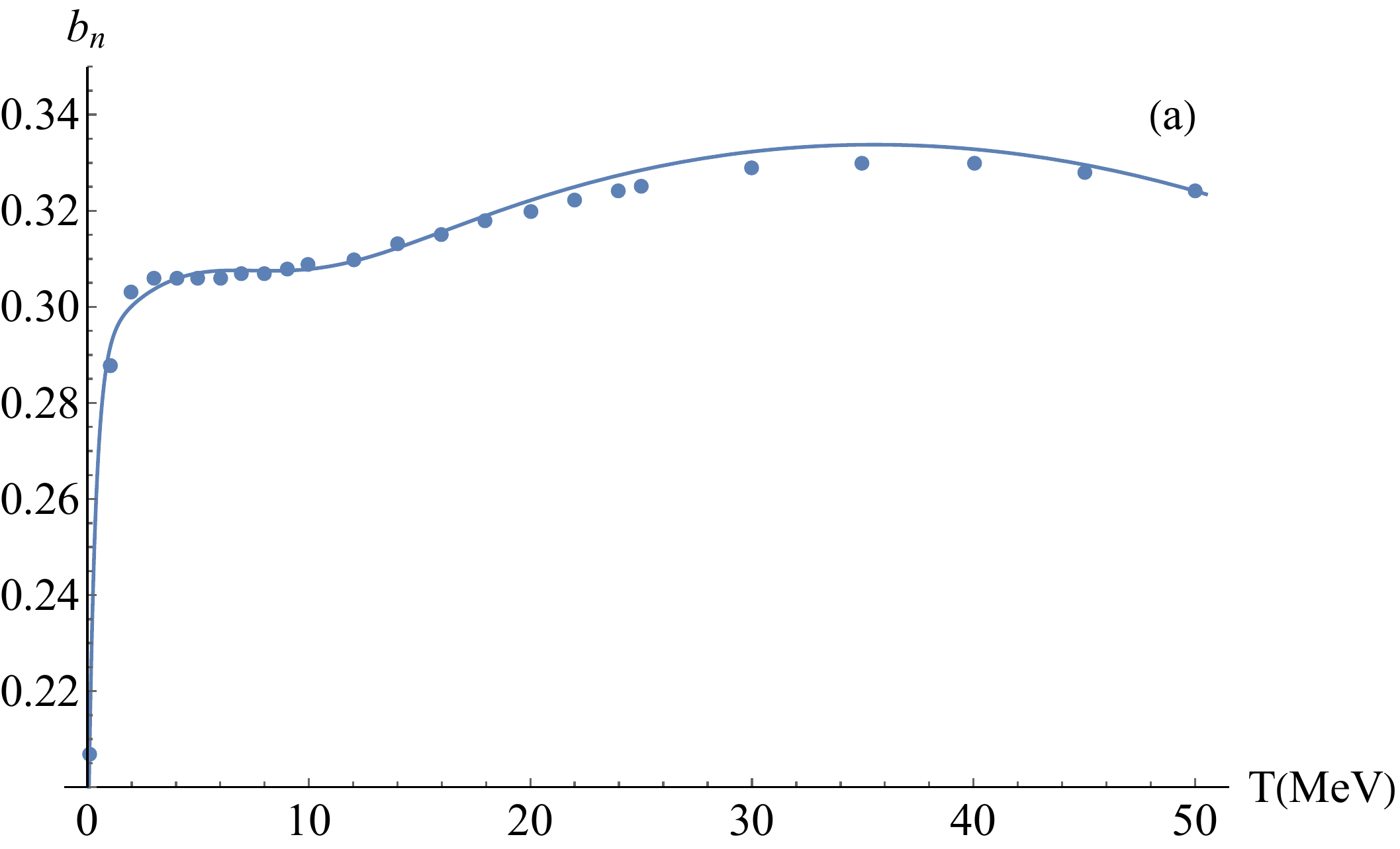}
  \includegraphics[width=3.1in]{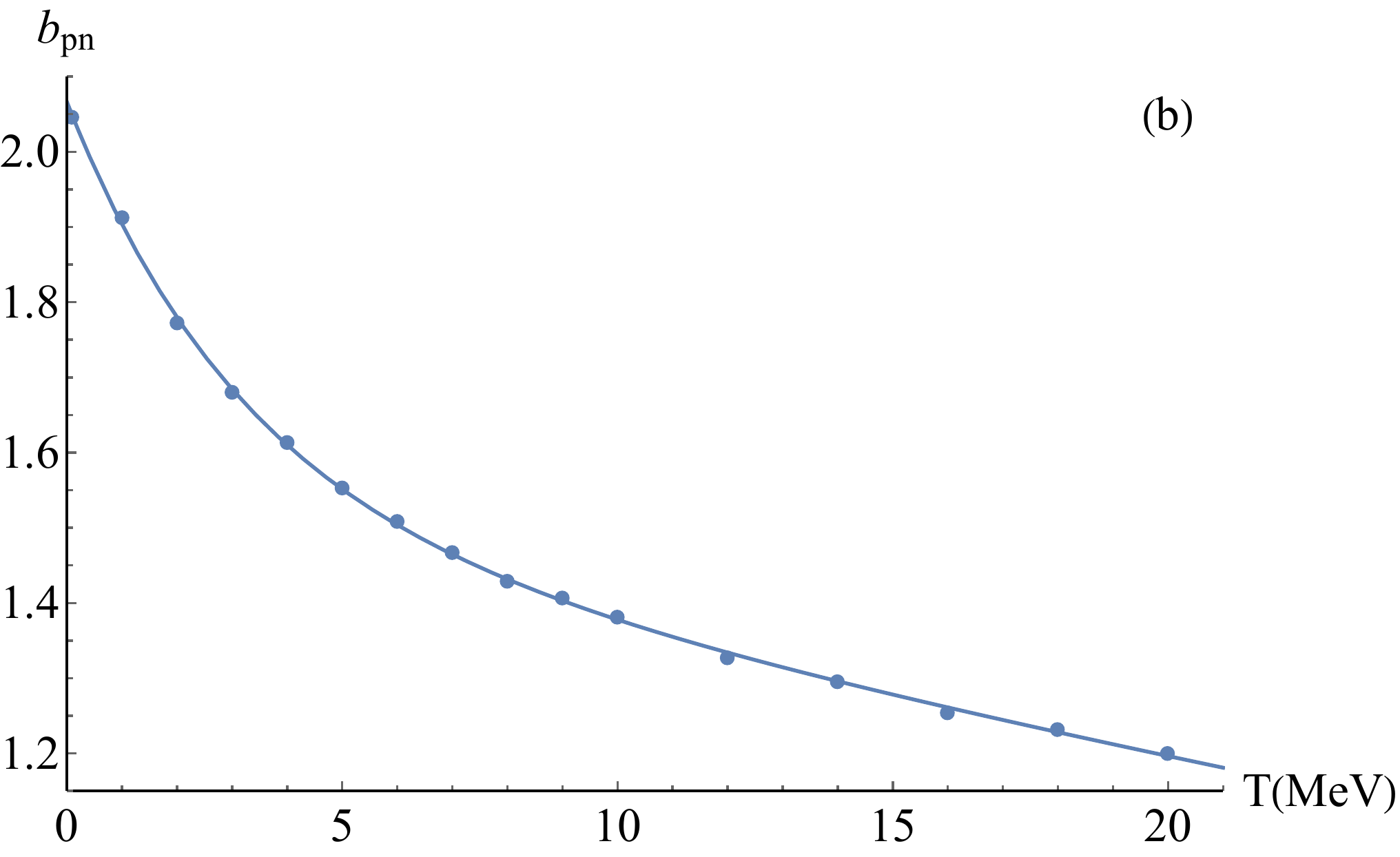}
  \caption{Fit of virial coefficients to data. The points at T=0.1 MeV
    are computed through effective range expansion.}
  \label{fig:vir}
\end{figure}

For the neutron matter virial coefficient, we use the data given in
Refs.~\cite{Horowitz06,Horowitz06b} and add three points at $T=0.1,
0.5$ and $150$ MeV, with virial coefficients of 0.207, 0.272, and
$2^{-5/2}$ respectively. The first two are determined from an
effective range expansion to the phase shift with scattering length
$-$18.9 fm and effective range 2.75 fm as determined from
Ref.~\cite{Gardestig09}. The last value at $T=150$ MeV is the
non-interacting result. We fit this data to a 10-parameter functional
form
\begin{eqnarray}
  b_n(T) &=&  b_0 + b_1 T + b_2 T^2 + b_3 T^3 +
  b_4 e^{-b_5 (T- b_6)^2} \nonumber \\
  &&+ b_7 e^{-b_8 (T- b_9)} \, .
\end{eqnarray}
We find that the parameter set $b_0=0.28745$,
$b_1=2.2006~\times~10^{-3}$ MeV$^{-1}$, $b_2= -2.6210 ~\times~10^{-5}$
MeV$^{-2}$, $b_3= 6.0617 ~\times~10^{-8} $ MeV$^{-3}$, $b_4= 1.0595
~\times~10^{-2}$, $b_5= 5.6734 ~\times~10^{-2}$ MeV$^{-2}$, $b_6=
3.4925$ MeV, $b_7=-2.7106 ~\times~10^{-3}$, $b_8= 3.1405$ MeV$^{-1}$,
$b_9=1.2010$ MeV matches the data. The data and the
fit are shown in the top panel of Fig.~\ref{fig:vir}. 

The contribution from the deuteron binding energy is typically
included in $b_{pn}(T)$, but in this work the deuteron binding energy
is removed (it will be added by the nuclear statistical equilibrium
part of the free energy in later work). For the low-temperature result
($T<1$ MeV), both $^1 S_0$ and $^3 S_1$ phase shifts contribute
at low energy, while due to the factor $e^{-E/2T}$
from Eq.\ (22) in Ref.~\cite{Horowitz06b}, it is reasonable to
ignore the higher-order phase shift contributions. We use the
scattering length $-$23.74 fm and effective range 2.77
fm~\cite{Gardestig09} for $^1 S_0$ channel and the scattering length
5.418 fm and effective range 1.833 fm for the $^3 S_1 $ channel from
Ref.~\cite{PavonValderrama:2005ku}. An alternate fit for $b_{pn}$ is
\begin{eqnarray}
  b_{pn}(T) = c_0 e^{-c_1(T+c_2)^2} + c_3 e^{-c_4 (T+ c_5)},
\end{eqnarray}
where $c_0=1.5273$, $c_1=1.7488 ~\times~10^{-4}$ MeV$^{-2}$,
$c_2=1.7550 ~\times~10^{1}$ MeV, $c_3= 0.45104 $, $c_4= 0.27513$
MeV$^{-1}$, $c_5= -1.1250$ MeV. The data and the fit are shown in the
bottom panel of Fig.~\ref{fig:vir}.

\subsection{First derivatives of the free energy}

From Eq.~(\ref{eq:fnp}), we can obtain the chemical
potentials and entropy directly
\begin{eqnarray}
  \mu_i & = &
  \mu_{\mathrm{i,vir}} g +
  f_{\mathrm{virial}}  \frac{\partial{g}}{\partial n_i} \nonumber \\
  && + \frac{\partial f_{\mathrm{deg}}}{\partial n_i}
  \left(1-g\right)-f_{\mathrm{deg}} 
  \frac{\partial g}{\partial n_i}
\end{eqnarray}
for $i=n,p$ 
where
\begin{equation}
  \mu_{\mathrm{i,vir}} \equiv
  \frac{\partial f_{\mathrm{virial}}}{\partial n_i} 
\end{equation}
and for the entropy
\begin{eqnarray}
  s &=& -\frac{\partial f_{\mathrm{virial}}}{\partial T}g-
  f_{\mathrm{virial}} \frac{\partial g}{\partial T}
  \nonumber \\ && -\frac{\partial{f_{\mathrm{deg}}}}{\partial T}
  \left(1-g \right)+f_{\mathrm{deg}} \frac{\partial g}{\partial T}
  \, .
\end{eqnarray}
In order to compute the derivatives of $g$ with respect to the
densities, one can differentiate Eqs.~(\ref{eq:virden}) with respect to
$n_n$ and $n_p$ and then solve the resulting four equations for the
quantities $\partial \mu_{i,\mathrm{vir}}/\partial n_j$ (for
$i,j=n,p$).

\subsection{Matter near nuclear saturation density}

Experimentally-measured nuclear masses are well-described by Skyrme
energy density functionals and thus it is expected that the energy
density of nuclear matter at zero temperature is also well-described
by the Skyrme model. In Ref.~\cite{Kortelainen14}, the parameters of
the Skyrme model were fit to several nuclear masses, charge radii, and
pairing energies using Bayesian inference. We use a set of 1000 Skyrme
parameterizations selected from the posterior distribution computed in
Ref.~\cite{Kortelainen14} to describe isospin symmetric matter.

Because there is relatively little information from theory or
experiment in some regimes, the Skyrme model will also be extrapolated
to higher densities and temperatures from below. This means that,
however, we cannot use all of the parameterizations because some of
them have a nucleon effective mass which becomes negative for
densities below $n_B<2~\mathrm{fm}^{-3}$. We remove such Skyrme models
from consideration.

Nuclear mass measurements are restricted to relatively
isospin-symmetric nuclei, thus neutron matter is not necessarily
accurately described by Skyrme models (see e.g. the discussion
regarding large fluctuations in the isovector channel in
Ref.~\cite{NavarroPerez18me}). Zero-temperature neutron matter up to
nuclear saturation density is tractable in quantum Monte
Carlo~\cite{Gandolfi12} and many-body perturbation
theory~\cite{Kruger13,Coraggio13}. It has thus become common to fit to
neutron matter calculations as well as nuclear mass
data~\cite{Chabanat95,Steiner05}. However, this practice presumes that
the Skyrme functional is well-suited to describing pure neutron
matter, an assumption that is not necessarily valid. Thus, for pure
neutron matter we use the four-parameter expression based on quantum
Monte Carlo results from Ref.~\cite{Gandolfi12},
\begin{eqnarray}
  &{\varepsilon}_{\mathrm{QMC}}(n_B) =
  f_{\mathrm{QMC}}(n_B) & \nonumber \\
  & = n_B\left[ a
    \left( \frac{n_B}{n_0}\right)^\alpha+ b \left(\frac{n_B}{n_0}
    \right)^\beta \right] & \, .
\end{eqnarray}
The range for the parameters $0.47 < a < 0.53$ and $12~\mathrm{MeV} <
\alpha < 13~\mathrm{MeV} $ is chosen as in Ref.~\cite{Steiner15un} to
enclose the limits in Ref.~\cite{Gandolfi12}.

The symmetry energy implied by many of the Skyrme fits, when combined
with the quantum Monte Carlo results for neutron matter, naturally
implies bound neutron matter at subsaturation densities. Similarly,
much of the range for $S$ and $L$ implied by the Skyrme
parameterizations is outside the allowed range from
Ref.~\cite{Tews17}. Thus we ignore the values for $S$ and $L$ from the
Skyrme models and limit $L$ between 44 and 65 MeV, and $S$ between
29.5 and 36.1 MeV as in Ref.~\cite{Steiner15un}. These bounds are
consistent with recent microscopic constraints \cite{Holt17} on the
density dependence of the symmetry energy from chiral effective field
theory. The prescription $(9.17\,S - 266~\mathrm{MeV}) < L < (14.3\,S
- 379~\mathrm{MeV})$ is used to ensure that $S$ and $L$ are
correlated. The coefficients $b$ and $\beta$ are determined by
\begin{eqnarray}
  b= S - a + \left(E/A\right)_{\mathrm{sky}}
\end{eqnarray}
\begin{eqnarray}
  \beta=\frac{1}{b} \left(\frac{L}{3} - \alpha a\right)
\end{eqnarray}
where $\left(E/A\right)_{\mathrm{sky}}$ is the binding energy per
particle of nuclear matter from the Skyrme interaction. Finally, we
combine the nuclear matter and neutron matter results by assuming the
symmetry energy is quadratic in $x_p$. This choice ensures that
nuclear matter is representative of experimental results on nuclear
masses while neutron matter agrees with modern theory results.

We note that the free energy density of matter from the QMC results
above is always at least linear in the density, and by
Eq.~(\ref{eq:virden}) at least linear in the fugacity at low densities.
This also holds for the Skyrme model, since the kinetic part of the
energy density is proportional to $k_F^5$ and the potential energy
part is proportional to at least one power of the density. Thus our
function $g$ in Eq.~(\ref{eq:g}) above is defined so that
$f_{\mathrm{deg}}$ will leave the second-order virial coefficients
unchanged from the values determined by experiment in the low-density
limit.

\subsection{Matter at high densities}

Above nuclear saturation density, there are two principal sources for
constraints on the EOS of matter: heavy ion collisions and neutron
star observations. Constraints on the EOS from heavy ion collisions
near the saturation density do not yet contradict results from Skyrme
fits. On the other hand, constraints from heavy ion collisions on the
EOS at higher densities do not yet provide a clear picture. Until the
results from heavy ion collisions are more definitive, the Skyrme
model from above is extrapolated to higher densities to describe
isospin-symmetric nuclear matter.

Neutron star mass and radius observations constrain the equation of
state of neutron-rich matter at high densities, in particular, the
pressure as a function of the energy density~\cite{Steiner10te}.
Unfortunately, neutron star observations do not yet currently
constrain the proton fraction of neutron star matter.
We find that the form
\begin{eqnarray}
  \varepsilon_{\mathrm{NS}}(n_B)
  = f_{\mathrm{NS}}(n_B) &=& p_0 n_B \sqrt{n_B} +
  p_1 n_B^2 + p_2 n_B^2 \sqrt{n_B} 
  \nonumber \\ && + p_3 n_B^3 + p_4 n_B^4 
  \label{eq:nsfit}
\end{eqnarray}
provides a good fit to the results from Ref.~\cite{Steiner10te}. We
randomly select EOSs from a Markov chain constructed in
Ref.~\cite{Steiner15un}, and fit them to Eq.~(\ref{eq:nsfit}).
Ref.~\cite{Steiner15un} constructed several Markov chains, and we use
the chain which was constructed using GCR (as above; from
Ref.~\cite{Gandolfi12}), Model A (which models high-density matter
using polytropes), and includes all of the mass and radius data from
photospheric radius expansion X-ray bursts and quiescent low-mass
X-ray binaries.

Astrophysical simulations can probe densities larger than those
constrained by the neutron star data in Ref.~\cite{Steiner15un}.
Between a transition density, $n_{Bf}$ and the largest
baryon density we consider, $n_B = 2~\mathrm{fm}^{-3}$, we
implement a simple EOS adapted from Constantinou and
Prakash~\cite[herafter denoted C\&P]{Constantinou17}.
The procedure for matching these EOSs begins by 
setting $n_{Bf}$ equal to the highest density specified
by the Monte Carlo data in Ref.~\cite{Steiner15un}. We decrease
this transition density as necessary to ensure that the EOS from
Eq.~(\ref{eq:nsfit}) is causal for densities lower than this
transition density. We add an additional parameter, $\phi$,
which is equal to the speed of sound at the largest density we
consider, $n_B = 2~\mathrm{fm}^{-3}$.

The EOS between $n_B=n_{Bf}$ and $n_B = 2~\mathrm{fm}^{-3}$ is chosen
depending on the relative magnitude of the speed of sound at these two
endpoints. If the speed of sound is increasing with increasing baryon
density, then, we choose
\begin{equation}
  c_s^2 = 1-a_1 +\frac{a_1 a_2 n_B^{a_1}}{1+a_2 n_B^{a_1}}
\end{equation}
and determine $a_1$ and $a_2$ by matching the
boundaries $\equiv c_s^2 \left(n_{B f}\right)$ and
$\phi=c_s^2 \left( n_B=2~\mathrm{fm}^{-3}\right)$ thus
ensuring $\beta \rightarrow 1$ as $n_B \rightarrow \infty $.
The energy density above $n_B = n_{Bf}$ is 
\begin{equation}
  \varepsilon_{NS} = -m_n n_B + c_1 \left( \frac{1}{2} a_2 n_B^2+ \frac{n_B^{2-a_1}}{2-a_1} \right) + c_2,
\end{equation}
where
\begin{eqnarray}
  c_1 &=& \frac{\varepsilon_f+m_n {n_{Bf}}+ P_f}{n_{Bf}^2
    \left(a_2+n_{Bf}^{-a_1}\right)}
  \\ \nonumber
  c_2 &=& \frac{1}{2} \left[\left(\varepsilon_f +m_n n_{Bf}
    -P_f \right)+a_1 \frac{\varepsilon_f +m_n n_{Bf}+P_f}
    {\left(a_1-2\right)\left(1+a_2 n_{Bf}^{a_1}\right)}\right]
\end{eqnarray}
Alternatively, if the speed of sound is decreasing with increasing
density, then we set
\begin{equation}
  c_s^2 = a_1 - \frac{a_1 a_2 n_B^{a_1}}{1+a_2 n_B^{a_1}}
\end{equation}
and match the boundaries as before
ensuring $\beta \rightarrow 0$ as $n_B \rightarrow 
\infty $. The corresponding energy density is 
\begin{equation}
  \varepsilon_{NS} = \frac{c_1 n_B \, {}_2 F_1 \left(1, -\frac{1}{a_1},
    1-\frac{1}{a_1}, -\frac{{n_B}^{-a_1}}{a_2}\right)}{a_2}+c_2 - m_n n_B
\end{equation}
where ${}_2 F_1$ is a hyper-geometric function with Pfaff's
transformation and the constants $c_1$ and $c_2$ are
\begin{eqnarray}
  c_1 &=& n_{B_f}^{-a_1-1} \left(a_2 n_{B_f}^{a_1}+1\right)
  \left(\varepsilon_f + m_n n_{B_f}
  + P_f \right)
  \\ \nonumber
  c_2 &=&  n_{B_f}^{-a_1} \left[a_2 n_{B_f}^{a_1} 
    \left(\varepsilon_f + m_n n_{B_f}
     \right) \right.
    \\ \nonumber
    && - \left(a_2 n_{B_f}^{a_1} +1\right) {}_2 F_1 \left(1, -\frac{1}{a_1},
    1-\frac{1}{a_1}, -\frac{n_B^{-a_1}}{a_2}\right)
    \\ \nonumber
    && \left. \left(\varepsilon_f + m_n n_{B_f} + 
    P_f \right) \right]
  \frac{1}{a_2} \, .
\end{eqnarray}
Although in practice $\phi$ is chosen randomly so this is
rare, if $c_s^2 \left(n_{B f}\right)=
\phi=c_s^2 \left( n_B=2~\mathrm{fm}^{-3}\right)$, then we ensure
$c_s^2$ is constant at high densities.
The corresponding energy density is 
\begin{eqnarray}
  \varepsilon_{NS} &=& -m_n n_B + \frac{\left(\varepsilon_f+
    m_n n_{B_f} + P_f\right)}{\left(1+C_{s_f}^2\right)}
  \left( {\frac{n_B}{n_{B_f}}}\right)^{1+C_{s_f}^2}
  \nonumber \\ 
  && + \frac{C_{s_f}^2 \left( \varepsilon_f +
    m_n n_{B_f}\right)-P_f}{1+C_{s_f}^2}.
\end{eqnarray}
This speed of sound correction ensures that neutron star matter is
causal, but an additional correction (described below) will be
required to ensure that the speed of sound is not larger than the
speed of light at all temperatures and electron fractions.

In order to combine information from QMC near the saturation
density and information from neutron star observations at
higher densities, we define a function $h$,
\begin{eqnarray}
  h = \frac{1}{1+\exp{\lbrack \gamma(n_B - \frac{3}{2} n_0 )}\rbrack}
\end{eqnarray}
where $\gamma$ is 20.0 fm$^{3}$. This function is
used to interpolate between the two density regimes.

\subsection{Hot matter near the saturation density}

Nuclear two- and three-body forces based on chiral effective theory have
shown great progress in computing the EOS of matter near nuclear
saturation. The Kohn-Luttinger-Ward perturbation series can be used to
compute the EOS of matter at finite temperature as described in
Refs.~\cite{Wellenhofer+14,Wellenhofer15to}. The resulting EOS can then be 
fitted with a Skyrme interaction, as done for example in Ref.~\cite{Lim17}.
However, it is difficult to use these results to 
quantify the uncertainties in these EOS calculations for matter at
$T=0$ where large cancellations between attractive and repulsive 
interactions lead to large theoretical errors.

To attempt to address this, we refit only the finite-temperature
correction from the chiral EOS,
\begin{equation}
  \Delta f_{\mathrm{hot}}(n_B,x_p,T) \equiv
  f_{\mathrm{hot}}(n_B,x_p,T) - f_{\mathrm{hot}}(n_B,x_p,T=0)
\end{equation}
and add these finite temperature corrections on top of our EOS. The
EOS for neutron matter ($x_p=0$) and nuclear matter ($x_p=1/2$) is
obtained from the perturbation series (including the non-interacting
contribution) and fitted with a single Skyrme model. The resulting
parameter set is given in Table~\ref{tab:skhot}. We assume that these
finite temperature corrections are quadratic in the isospin asymmetry,
$\delta$. The EOS is not fully
quadratic~\cite{Steiner06hs,Wellenhofer16}, but the quadratic
approximation is good enough in comparison to the uncertainties in the
nature of the strong interaction above the saturation density. We do
not expect this Skyrme interaction to give a reasonable decription of
nuclei or saturated nuclear matter, because we only employ it to
describe the finite temperature part of the EOS. There are some
regions, especially at large densities, for which the EOS is unstable,
i.e. $ds/dT < 0$, but these regions most often result in an acausal
EOS and are thus fixed by the speed of sound correction described
below.

\begin{table}
  \begin{tabular}{cl}
    parameter & value \\
    \hline
    $x_0$ & $\hphantom{-}4.19756~\times 10^{1}$\\
    $x_1$ & $-6.94792~\times 10^{-2}$\\
    $x_2$ & $\hphantom{-}4.19202~\times 10^{-1}$\\
    $x_3$ & $-2.87797~\times 10^{1}$\\
    $t_0$ & $\hphantom{-}5.06729~\times 10^{3}~\mathrm{fm}^{2}$\\
    $t_1$ & $\hphantom{-}1.74925~\mathrm{fm}^{4}$\\
    $t_2$ & $-4.72119~\times 10^{-1}~\mathrm{fm}^{4}$\\
    $t_3$ & $-1.94596~\times 10^{5}~\mathrm{fm}^{2+3\alpha}$\\
    $\alpha$ & $\hphantom{-}1.44165~\times 10^{-1}$\\
  \end{tabular}
  \caption{Skyrme parameters obtained from the chiral EOS used for the
  finite-temperature corrections in this work}
  \label{tab:skhot}
\end{table}

\subsection{The full combined EOS}

First, we define the symmetry energy to
include a zero-temperature contribution which combines
the QMC EOS near saturation density, the neutron star fit at higher
densities, and the Skyrme interaction for isospin-symmetric matter
\begin{eqnarray}
  {\varepsilon}_{\mathrm{sym}}(n_B) &=&
  h (n_B) {\varepsilon}_{\mathrm{QMC}}(n_B) +
  \left[ 1-h(n_B) \right] \varepsilon_{\mathrm{NS}}(n_B)
  \nonumber \\
  && - f_{\mathrm{Skyrme}}(n_B,x_p=1/2,T=0) \, .
\end{eqnarray}
Defining the isospin asymmetry $\delta=1-2 x_p$, we can combine this
with the model described above to obtain the free energy density of
degenerate matter
\begin{widetext}
  \begin{eqnarray}
    f_{\mathrm{deg}}(n_B,x_p,T) &=&
    f_{\mathrm{Skyrme}}(n_B,x_p=1/2,T=0)
    + \delta^2 {\varepsilon}_{\mathrm{sym}}(n_B)
    + \nonumber \\
    && + \delta^2 \Delta f_{\mathrm{hot}}(n_B,x_p=0,T)
    + (1-\delta^2) \Delta f_{\mathrm{hot}}(n_B,x_p=1/2,T).
  \end{eqnarray}
\end{widetext}
Finally, we ensure that the total nucleonic
free energy gives the result from the virial expansion at
high temperatures using Eq.~(\ref{eq:fnp}).
When we need to include the electrons, positrons,
and photons, we define the free energy density
\begin{equation}
  f_{npe\gamma} \equiv f_{\mathrm{np}} +f_{e^-}+f_{e^+}+f_{\gamma} \, .
\end{equation} 

Using this formalism, the chemical potentials and entropy
can be computed directly:
\begin{eqnarray}
  \frac{\partial f_{\mathrm{deg}}}{\partial n_n}  &=&
 \frac{1}{2} \mu_{n,\mathrm{Skyrme}}(n_B,x_p=1/2,T=0)
 \\ \nonumber
 &&+ \frac{1}{2} \mu_{p,\mathrm{Skyrme}}(n_B,x_p=1/2,T=0)
  \nonumber \\ && + \delta^2
  \frac{\partial \varepsilon_{\mathrm{sym}}}{\partial n_B} +
\frac{2 \delta (1-\delta)}{n_B} \varepsilon_{\mathrm{sym}}
  \nonumber \\ &&
  + \frac{2 \delta (1-\delta)}{n_B}
\Delta f_{\mathrm{hot}}(n_B,x_p=0,T)
  \nonumber \\
  &&  + \delta^2 \Delta \mu_{n,\mathrm{hot}}(n_B,x_p=0,T) 
  \nonumber \\ &&
  - \frac{2 \delta (1-\delta)}{n_B}
  \Delta f_{\mathrm{hot}}(n_B,x_p=1/2,T) 
  \nonumber \\
  &&  +  \left( 1-\delta^2 \right) \Delta \mu_{n,\mathrm{hot}}
    (n_B,x_p=1/2,T),
\end{eqnarray}
\begin{eqnarray}
  \frac{\partial f_{\mathrm{deg}}}{\partial n_p}  &=&
 \frac{1}{2} \mu_{p,\mathrm{Skyrme}}(n_B,x_p=1/2,T=0)
  \nonumber \\ &&  +\frac{1}{2} \mu_{n,\mathrm{Skyrme}}(n_B,x_p=1/2,T=0)
  \nonumber \\ && + \delta^2
  \frac{\partial \varepsilon_{\mathrm{sym}}}{\partial n_B} -
  \frac{2 \delta \left( 1 + \delta \right)}{n_B} \varepsilon_{\mathrm{sym}}
  \nonumber \\ &&
  - \frac{2 \delta \left( 1 + \delta \right)}{n_B}
  \Delta f_{\mathrm{hot}}(n_B,x_p=0,T) 
  \nonumber \\
  &&  + \delta^2 \Delta \mu_{p,\mathrm{hot}}(n_B,x_p=0,T) 
  \nonumber \\ &&
  + \frac{2 \delta \left( 1 + \delta \right)}{n_B}
  \Delta f_{\mathrm{hot}}(n_B,x_p=1/2,T) 
  \nonumber \\
  &&  + \left( 1-\delta^2 \right) \Delta \mu_{p,\mathrm{hot}}
    (n_B,x_p=1/2,T),
\end{eqnarray}
and
\begin{eqnarray}
  \frac{\partial f_{\mathrm{deg}}}{\partial T}  &=&
  -  \delta^2 s_{\mathrm{hot}} \left(n_B, x_p=0,T \right) - 
  \nonumber \\
  &&  \left( 1-\delta^2 \right) 
  s_{\mathrm{hot}} \left(n_B, x_p=1/2,T \right),
\end{eqnarray}
where
\begin{eqnarray}
  \frac{\partial \varepsilon_{\mathrm{sym}}}{\partial n_B} &=&
  h^{\prime}(n_B) {\varepsilon}_{\mathrm{QMC}}(n_B)
  + h(n_B) {\varepsilon}^{\prime}_{\mathrm{QMC}}(n_B)
  \nonumber \\ &&
  - h^{\prime}(n_B) {\varepsilon}_{\mathrm{NS}}(n_B)
  + \left[1-h(n_B)\right] {\varepsilon}^{\prime}_{\mathrm{NS}}(n_B) +
  \nonumber \\ &&
  - \frac{1}{2} \left[\mu_{n,\mathrm{Skyrme}}(n_B,x_p=1/2,T=0) \right.
    \nonumber \\ &&
    + \left. \mu_{p,\mathrm{Skyrme}}(n_B,x_p=1/2,T=0) \right].
\end{eqnarray}

In summary, we have 5 parameters: (i,ii) the values of $a$ and
$\alpha$ which determine sub-saturation neutron matter, (iii,iv) the
values of $S$ and $L$ which determine the symmetry energy and its
density dependence, and (v) the value, $\phi$, of the speed of sound
in neutron star matter at $n_B=2~\mathrm{fm}^{-3}$. In addition, we
have two indexes which enumerate random samples from posterior
distributions including (i) the index of the Skyrme parameterization
from Ref.~\cite{Kortelainen14} and (vi) the index of the neutron star
EOS from the Markov chain generated in Ref.~\cite{Steiner15un}.

\subsection{Enforcing causality at high density}
\label{s:causal}

Since our phenomenological EOS does not have manifest Lorentz
covariance, it has the potential to become acausal at high-densities.
At every electron fraction and temperature, there may be a baryon
density, $n_B^{*}$, above which the EOS becomes acausal.
Because our phenomenological EOS (as all other EOS tables) operate
as functions of the densities and temperatures, it is useful to
rewrite the speed of sound in terms of derivatives of the
Helmholtz free energy. This is done in Appendix I below for a
general system with any number of conserved charges (though here we
only have two, baryon number and electric charge). 

When our phenomenological EOS becomes acausal above some baryon
density, $n_B^{*}$, we replace the EOS with a causal EOS,
$\tilde{f}_{C\&P}$, following the prescription in
Ref.\ \cite{Constantinou17}. We construct a modified free energy
density with the following
\begin{equation}
  \tilde{f}_{all}=\tilde{f}_{npe\gamma} \Theta(n_{B}^{*}-n_B)+
  \tilde{f}_{C\&P} \Theta(n_B-n_{B}^{*}),
\end{equation}
where contributions from electrons, positrons and photons are included
in $\tilde{f}_{npe\gamma}$. To be more concise, we suppress the
subscripts $npe\gamma$ in the following. Using $\varepsilon$ for
energy density (including rest mass energy density), $S$ for entropy,
$s$ for entropy density, and $\tilde{s}$ for entropy per baryon, the
C\&P speed of sound is
\begin{equation}
c_s^2 = \left( \frac{dP}{d\varepsilon} \right)_{\tilde{s},N_B,N_e}
= \left( \frac{dP}{d\varepsilon} \right)_{\tilde{s},N_B,Y_e}.
\end{equation}
Note that $\tilde{s}=S/N_B = s/n_B$, where $N_B$ is the number of
baryons. The C\&P derivation begins by noting that
\begin{equation}
P = - \varepsilon + 
n_B \left(\frac{\partial \varepsilon}{\partial
	n_B}\right)_{\tilde{s},N_B,Y_e} \, .
\label{eq:cp1} 
\end{equation}
To see this we can write
\begin{eqnarray}
\left(\frac{\partial \varepsilon}{\partial
	n_B}\right)_{\tilde{s},N_B,Y_e}  &&=
\left[\frac{\partial (E/V)}{\partial
	V}\right]_{S,N_B,Y_e}
\left[\frac{\partial (N_B/V)}{\partial
	V}\right]_{S,N_B,Y_e}^{-1} 
\nonumber \\
&&\hspace{-.3in}=\left(-\frac{P}{V} - \frac{E}{V^2}\right)
\left(-\frac{N_B}{V^2}\right)^{-1} = \frac{(P+\varepsilon)}{n_B}.
\end{eqnarray}
Taking the derivative of Eq.~(\ref{eq:cp1}), we can also rewrite the
pressure as a second derivative
\begin{eqnarray}
\left(\frac{\partial P}{\partial n_B}\right)_{\tilde{s},N_B,Y_e} &=&
-\left(\frac{\partial \varepsilon}{\partial
	n_B}\right)_{\tilde{s},N_B,Y_e} +
\left(\frac{\partial \varepsilon}{\partial
	n_B}\right)_{\tilde{s},N_B,Y_e} 
\nonumber \\ \nonumber
&&+ n_B
\left(\frac{\partial^2 \varepsilon}{\partial
	n_B^2}\right)_{\tilde{s},N_B,Y_e} 
\\
&=&n_B \left( \frac{\partial^2 \varepsilon}{\partial n_B^2}
\right)_{\tilde{s},N_B,Y_e}.  
\end{eqnarray}
Thus we can proceed as C\&P do,
\begin{equation}
\left( \frac{\partial^2 \varepsilon}{\partial n_B^2}
\right)_{\tilde{s},N_B,Y_e} - \frac{c_s^2}{n_B}
\left( \frac{\partial \varepsilon}{\partial n_B}
\right)_{\tilde{s},N_B,Y_e} = 0.
\end{equation}

Following the analytical continuation, at every value of $\tilde{s}$,
$N_B$, and $Y_e$ in the acausal region, we can use the C\&P solution
\begin{eqnarray} \label{eq:cp}
\varepsilon_{C\&P}(\tilde{s},n_B,N_B,Y_e) &&=
\left[\frac{\varepsilon^{*}(\tilde{s},N_B,Y_e)+P^{*}(\tilde{s},N_B,Y_e)}
{\beta+1}\right]
\nonumber \\ \nonumber
&& \hspace{-.4in}\times \left[ \frac{n_B}{n_B^{*}
    (\tilde{s},N_B,N_e)} \right]^{\beta+1} 
\\
&& \hspace{-.4in}+\left[
\frac{\beta \varepsilon^{*}(\tilde{s},N_B,Y_e) - P^{*}(\tilde{s},N_B,Y_e) }
{\beta+1}\right] \, .
\end{eqnarray}
If we assume that $\varepsilon^{*}$, $P^{*}$, and $n_B^{*}$ are
volume independent, then since they are determined at a fixed
value of $n_B$,
they cannot separately depend on $N_B$. Thus the full energy density
also does not depend on $N_B$.
To show this explicitly, we start from
\begin{eqnarray} \label{eq:dE}
E&=&\mu_B N_B+\mu_L N_B Y_e +T S - P V,
\\ \nonumber
dE&=& \left( \mu_B+\mu_L Y_e +T \tilde{s} \right) dN_B +\mu_L N_B dY_e
\\ \nonumber
&&+T N_B d\tilde{s} -P dV.
\end{eqnarray}
where $\mu_L \equiv \mu_p + \mu_e - \mu_n$. On the right hand side,
\begin{equation}
dV=\frac{1}{n_B}d{N_B}-\frac{N_B}{{n_B}^2} d n_B
\end{equation}
and on the left hand side,
\begin{equation}
dE= d\left(\varepsilon \frac{N_B}{n_B} \right)
=\frac{N_B}{n_B} d\varepsilon+
\frac{\varepsilon}{n_B} dN_B -\frac{\varepsilon N_B}{{n_B}^2} dn_B.
\end{equation}
Substituting the above two equations back into Eq.~(\ref{eq:dE}),
we find
\begin{equation}
d{\varepsilon}= \mu_L n_B dY_e
+T n_B d\tilde{s} + \left(\mu_B + \mu_L Y_e + T \tilde{s} \right) d n_B
\end{equation}
and the $N_B$ dependence disappears. Therefore 
$\varepsilon_{C\&P}(\tilde{s},n_B,N_B,Y_e)=
\varepsilon_{C\&P}(\tilde{s},n_B, Y_e)$:
\begin{eqnarray}
\varepsilon_{C\&P}(\tilde{s},n_B,Y_e) &=&
\left[\frac{\varepsilon^{*}(\tilde{s},Y_e)+P^{*}(\tilde{s},Y_e)}
{\beta+1}\right]
\left[ \frac{n_B}{n_B^{*}(\tilde{s},Y_e)} \right]^{\beta+1} 
\nonumber \\
&& +\left[
\frac{\beta \varepsilon^{*}(\tilde{s},Y_e) - P^{*}(\tilde{s},Y_e) }
{\beta+1}\right] \, .
\end{eqnarray}
We also find 
\begin{equation}\label{eq:mul}
  \left( \frac{\partial \varepsilon_{C\&P}}{\partial Y_e}
  \right)_{\tilde{s},n_B}=\mu_L n_B,
\end{equation}
\begin{equation}\label{eq:T}
  \left( \frac{\partial \varepsilon_{C\&P}}{\partial \tilde{s}}
  \right)_{n_B,Y_e}=T n_B,
\end{equation}
\begin{equation}\label{eq:mub}
  \left( \frac{\partial \varepsilon_{C\&P}}{\partial n_B}
  \right)_{\tilde{s},Y_e}=\mu_B+\mu_L Y_e + T \tilde{s}.
\end{equation}
We choose to use this solution above the value of $n_B$, denoted
$n_B^{*}(\tilde{s},Y_e)$, for which $c_s^2 = \beta$.
From Eq.~(\ref{eq:T}) the temperature is 
\begin{equation}
T = \frac{1}{n_B}
\left(\frac{\partial \varepsilon_{C\&P}}{\partial
	\tilde{s}}\right)_{n_B,Y_e}.
\end{equation}
Thus computing the temperature using the method in
Ref.~\cite{Constantinou17} requires evaluating derivatives of the form
$(\partial n_B^{*}/\partial \tilde{s})_{Y_e}$. These derivatives are
computed along the surface for which the speed of sound is equal to
its largest possible value (which we set to 0.9). Because the speed of
sound requires two derivatives of the free energy, these derivatives
(derivatives along a line of constant $c_s^2$) require third
derivatives of the free energy. The chemical potentials require a
similar calculation based on Eqs.~(\ref{eq:mul}) and (\ref{eq:mub}) 
which we do not explicitly show here.

The numerical evaluation of third derivatives introduces quite a bit of
noise. Our phenomenological formalism ensures that these derivatives
can be computed analytically. We leave this calculation to future
work. In the mean time, we can more easily compute the temperature
by an indirect approach. Presuming we would like to compute the
EOS at some fixed values of the baryon density $\widehat{n_B}$,
electron fraction $\widehat{Y_e}$,
and temperature $\widehat{T}$, one needs to solve the five
equations
\begin{eqnarray}
c_{s,\mathrm{DS}}^2(n_{B1}^{*},\widehat{Y_e},{T_1}^{*}) &=& \beta \, , 
\nonumber \\
c_{s,\mathrm{DS}}^2(n_{B2}^{*},\widehat{Y_e},{T_2}^{*}) &=& \beta \, ,
\nonumber \\
\tilde{s}_{\mathrm{DS}} \left( n_{B1}^{*},\widehat{Y_e},
      {T_1}^{*}\right) &=& \tilde{s} \, ,
\nonumber \\
\tilde{s}_{\mathrm{DS}} \left( n_{B2}^{*},\widehat{Y_e},
      {T_2}^{*}\right) &=& \tilde{s} + \delta \tilde{s} \, ,
 \nonumber \\
\frac{1}{\widehat{n_{B}}} \frac{
  \epsilon_{\mathrm{C\&P}} (\tilde{s} + \delta \tilde{s},
  \widehat{Y_e},n_B)
  - \epsilon_{\mathrm{C\&P}} (\tilde{s}, \widehat{Y_e},n_B)}
     {\delta \tilde{s}}&=& \widehat{T}
     \label{eq:five}
\end{eqnarray}
for the five values $n_{B1}^{*}, n_{B2}^{*}, T_1^{*}, T_2^{*}$, and
$\tilde{s}$ given some small fixed stepsize in the entropy per baryon,
$\delta \tilde{s}$. Here ``DS'' denotes our EOS while ``C\&P'' denotes
Eq.~(\ref{eq:cp}) from Ref.~\cite{Constantinou17}. The required
numerical derivatives to compute the temperature in the last equation
of Eq.~(\ref{eq:five}) can then be computed directly from finite
differences. By solving the five equations above, we get
$\tilde{s}_{C\&P}\left( \widehat{n_B},\widehat{Y_e}, \widehat{T}
\right)$, and then
\begin{equation}
\tilde{f}_{C\&P}\left(\widehat{n_B},\widehat{Y_e}, \widehat{T}\right) =
\varepsilon_{C\&P} - \widehat{T}\widehat{n_B}\tilde{s}_{C\&P}
\left( \widehat{n_B},\widehat{Y_e}, \widehat{T} \right).
\end{equation}

\section{Results}

\begin{figure}
  \includegraphics[width=3.1in]{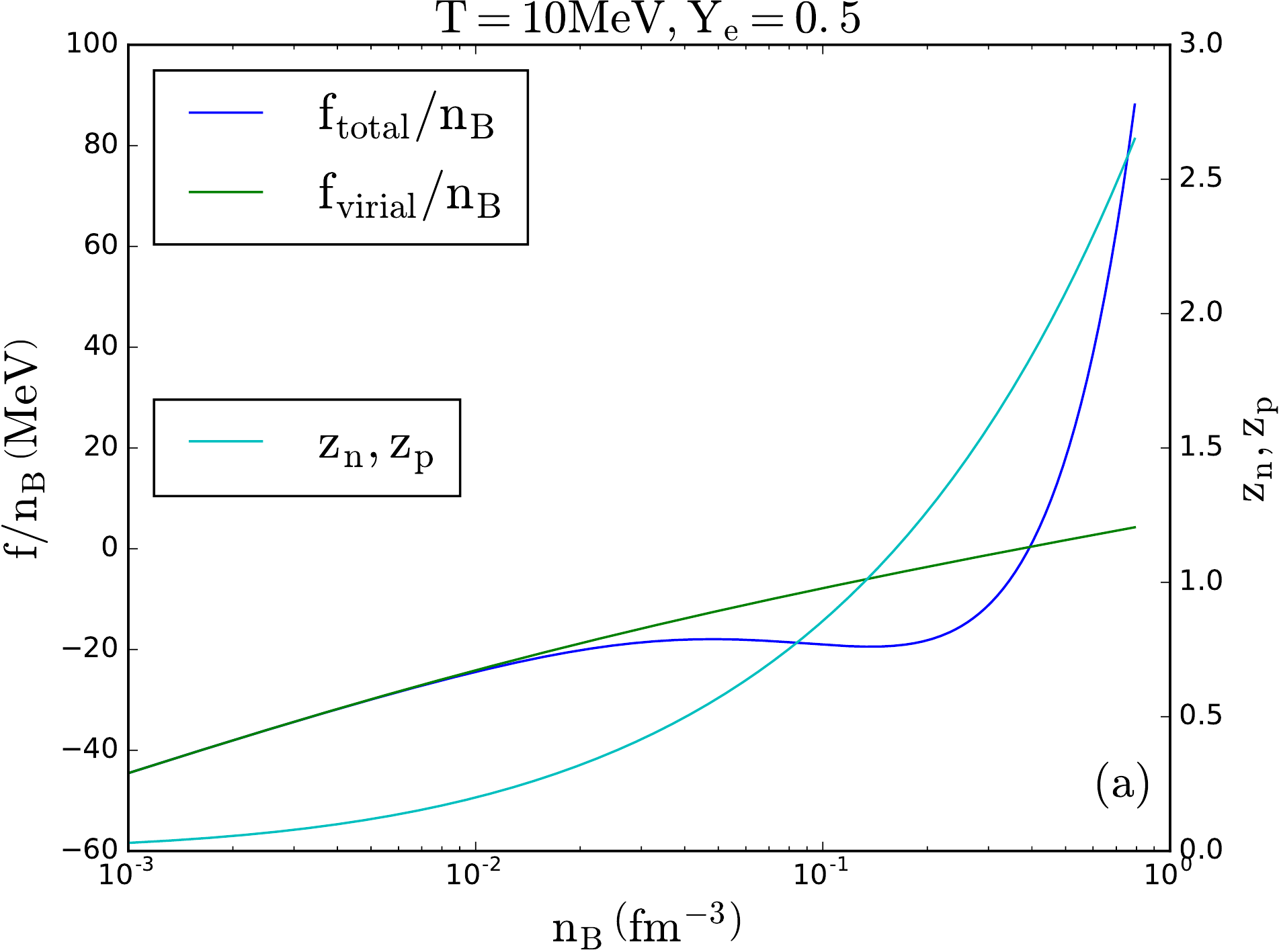}
  \includegraphics[width=3.1in]{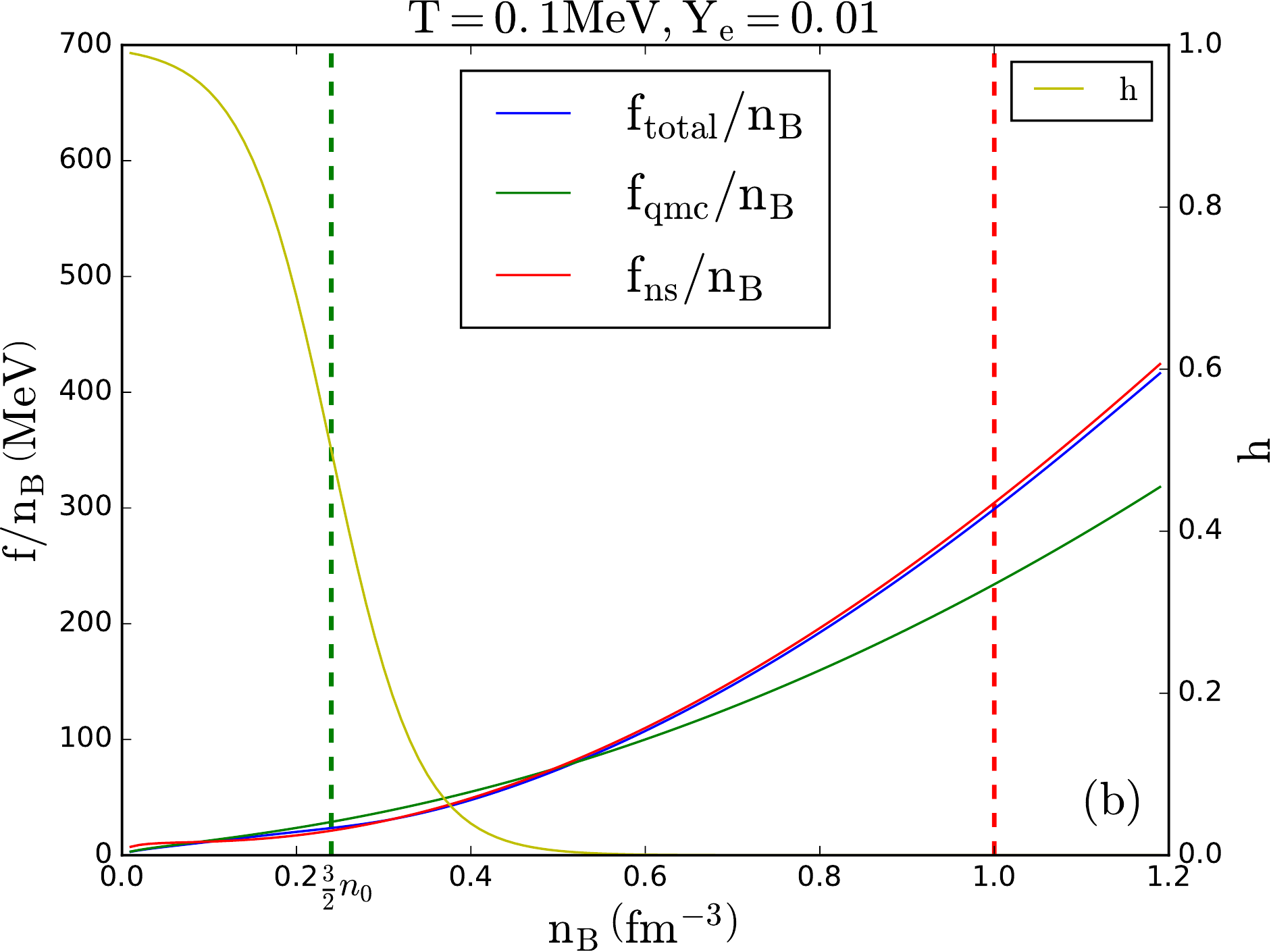}
  \includegraphics[width=3.1in]{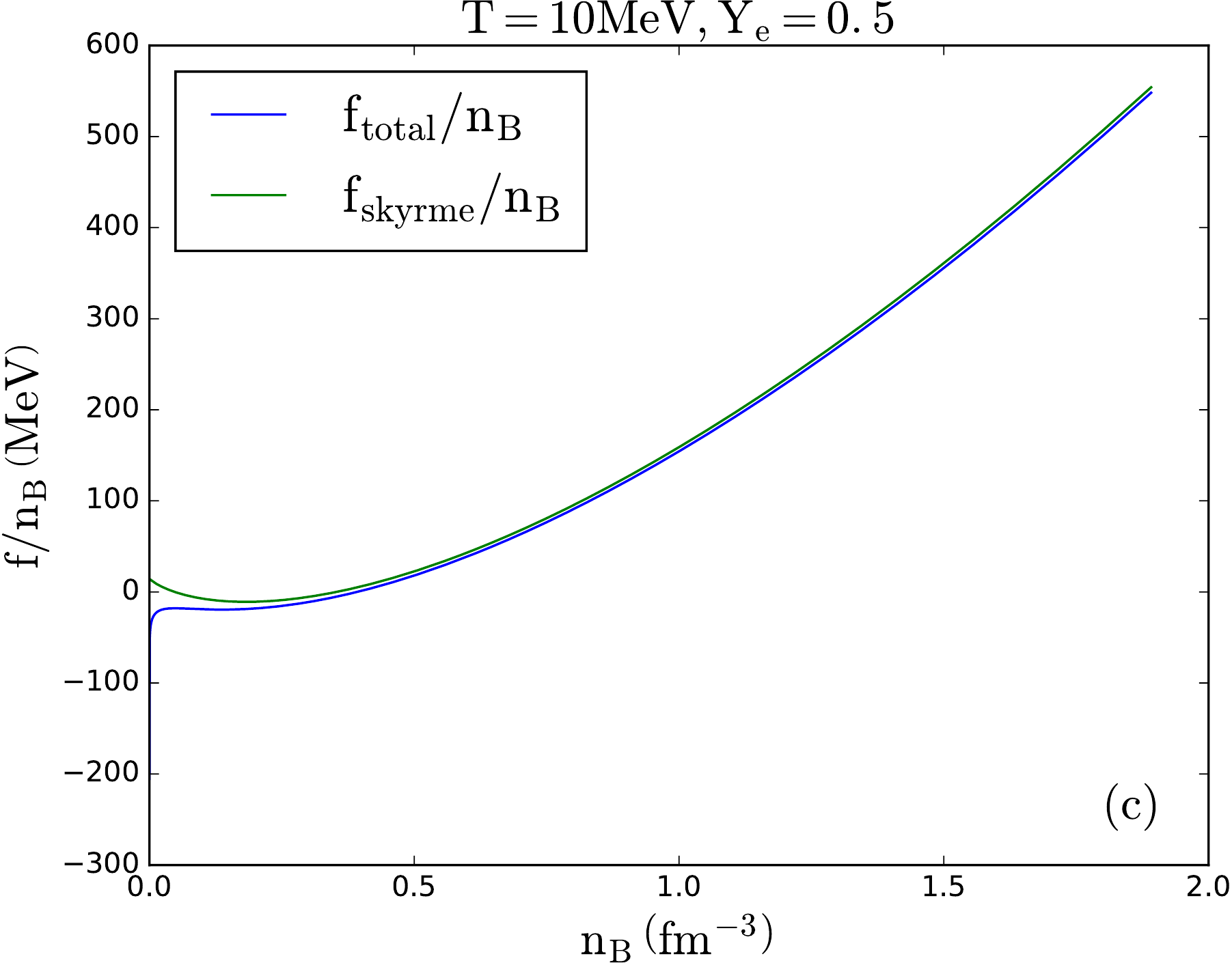}
  \caption{Three figures which show how the full EOS compares to the
    limiting forms. The top panel compares the full free energy to the
    result from the virial expansion and shows that they match at
    lower densities where the fugacities are much smaller than 1. The
    middle panel shows that the result for neutron matter matches the
    QMC free energy at low densities, the neutron star free energy at
    densities reached in the neutron star interiors, and is softened
    at high densities to ensure a speed of sound less than the speed
    of light. The bottom panel shows that free energy starts to
    deviate from the Skyrme interaction at higher temperatures as the
    corrections from the chiral EOS begin to contribute. These panels
    show the result from one of many EOSs generated in this work.}
  \label{fig:comp}
\end{figure}

Fig.~\ref{fig:comp} shows how our full EOS behaves in the (i)
non-degenerate limit, (ii) the limit of zero-temperature neutron
matter, and (iii) the limit of high-temperature and high-density
nuclear matter. Only one parameter set is chosen and the same
parameter set is chosen for each of the three panels. In the
non-degenerate limit, our full EOS smoothly matches on to the virial
EOS as determined by Eq.~(\ref{eq:g}). The middle panel shows that, in
the limit of zero-temperature neutron matter, our full EOS matches the
QMC result at low density and remains close to the neutron star EOS at
moderate densities. At higher densities, the free energy does not
increase too quickly with density because our correction for causality
begins to start becoming important. The bottom panel compares
our full EOS (for this parameterization) with the ($T=0$) Skyrme EOS,
showing a small modification in the EOS due to the finite temperature
correction from the chiral EOS.

We can construct a figure similar to Fig.~\ref{fig:comp} for any
physical combination of our model parameters: $\alpha$, $a$, $S$, $L$,
$\phi$ and Skyrme model from Ref.~\cite{Kortelainen14} (which
represents a 12-dimensional space of possible Skyrme models) and any
neutron star model from Ref.~\cite{Steiner15un} (which represents a
6-dimensional space of high-density EOSs).

\begin{figure}
  \includegraphics[width=0.5\textwidth]{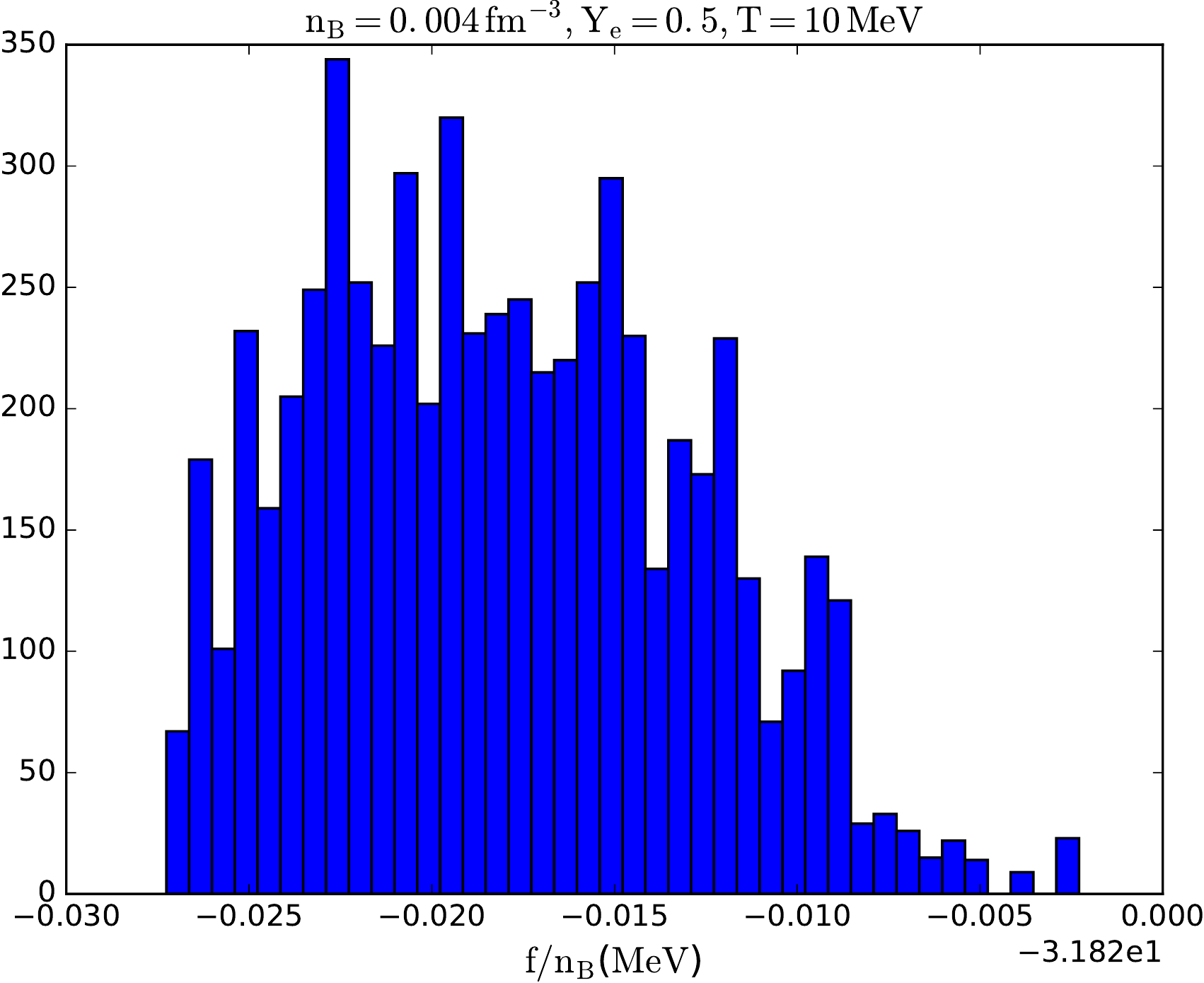}
  \caption{The probability distribution for the free energy per baryon
    at $n_B=0.004~\mathrm{fm}^{-3}$, $Y_e=0.5$, and $T=10~\mathrm{MeV}$.}
  \label{fig:mc_lowdens}
\end{figure}

For any baryon density, electron fraction, and temperature, we can
compute a probability distribution for the free energy per baryon over
our entire parameter space. The magnitude and shape of the uncertainty
in the free energy changes depending on the relevant physics in that
region which our model describes. Fig.~\ref{fig:mc_lowdens} (??The
titles on Figs.\ 3-7 should have $N_B \rightarrow n_B$, and I would
remove the ``/197.327'' in the temperature??) shows the variation in
the free energy in nondegenerate matter where the EOS is dominated by
the virial expansion. The small remaining uncertainty here originates
in the Skyrme model that is selected, and the non-smooth nature in the
probability distribution is a relic of the limited sampling size of
Skyrme interactions.

\begin{figure}
  \includegraphics[width=0.5\textwidth]{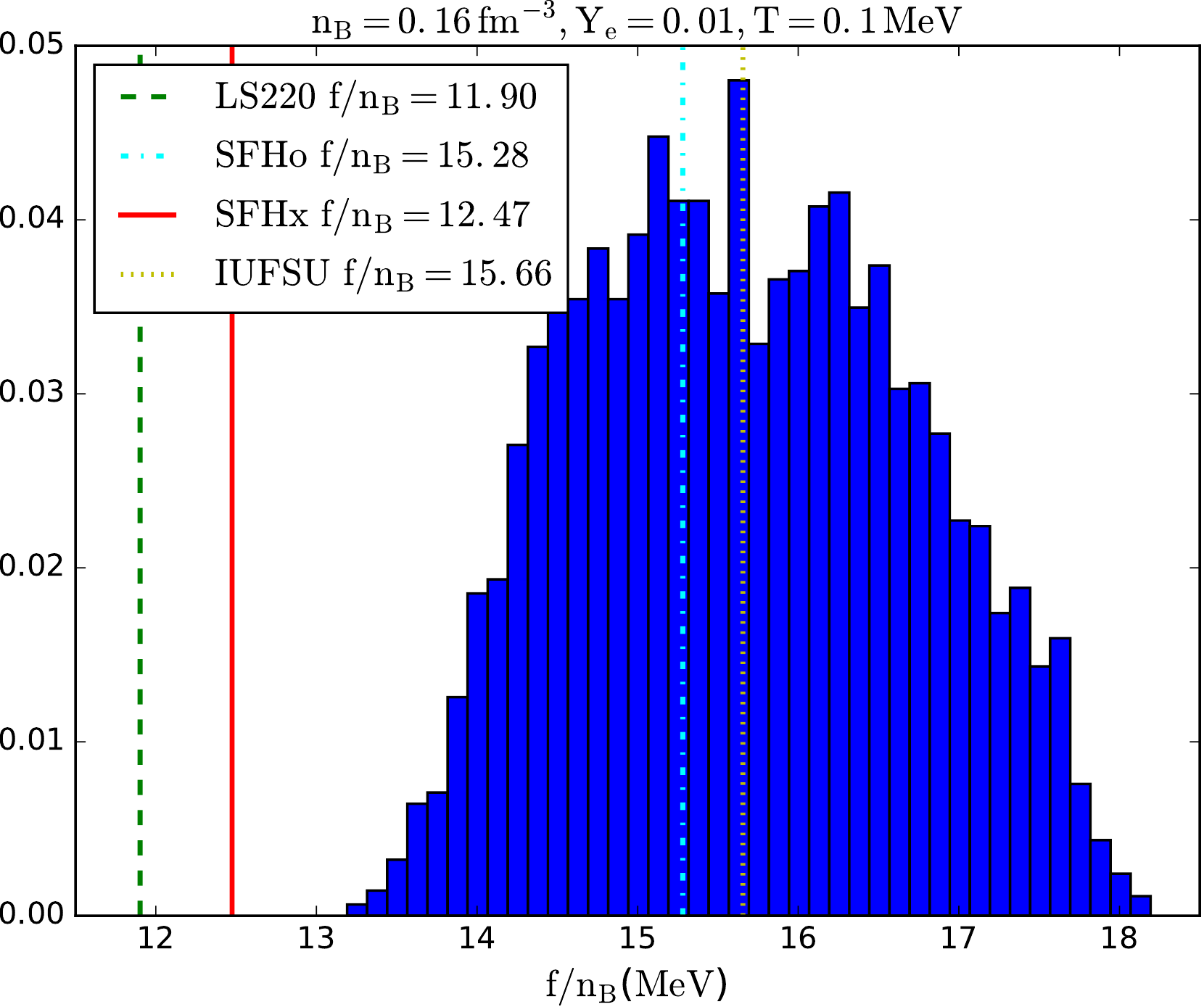}
  \caption{The probability distribution for the free energy per baryon
    at $n_B=0.16~\mathrm{fm}^{-3}$, $Y_e=0.01$, and
    $T=0.1~\mathrm{MeV}$.}
  \label{fig:mc_neut_sat}
\end{figure}

Fig.~\ref{fig:mc_neut_sat} shows the variation in the free energy per
baryon at nuclear saturation density in nearly pure neutron matter and
in the limit of zero temperature. The results for LS220, SFHo, SFHx,
and IUFSU are also shown. The distribution is centered around 16 MeV,
corresponding to a symmetry energy of 32 MeV, and values lower than 13
MeV are excluded in our model because they correspond to symmetry
energies lower than 29.5 MeV. Symmetry energies this small seem to be
excluded from Quantum Monte Carlo and chiral effective field theory
calculations of pure neutron
matter~\cite{Gandolfi12,Holt17,Drischler17}. See also
Ref.~\cite{Tews17} for a more general result that gives a similar
lower limit for $S$ of $28$ MeV.

\begin{figure}
  \includegraphics[width=0.5\textwidth]{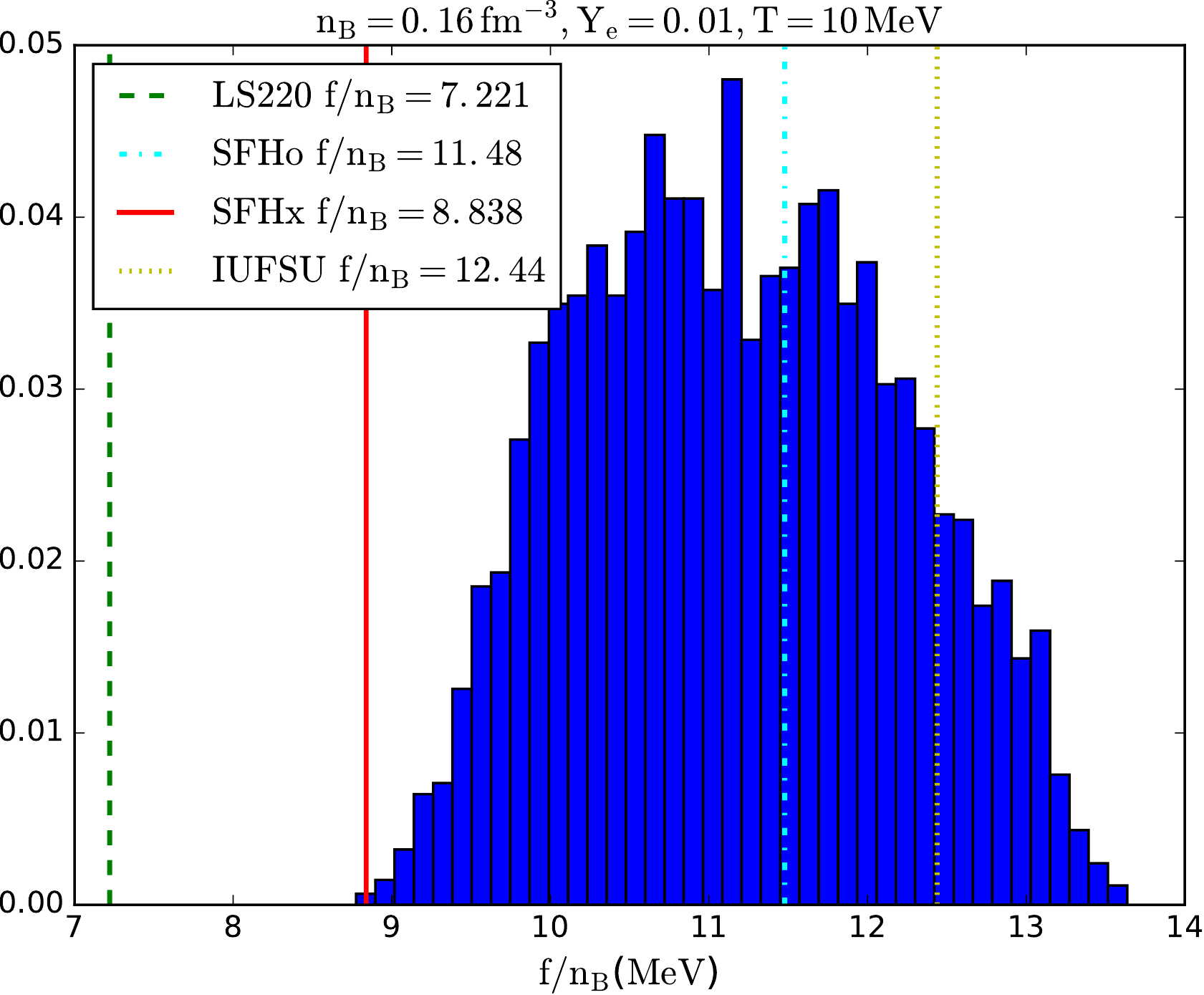}
  \caption{The probability distribution for the free energy per baryon
    at $n_B=0.16~\mathrm{fm}^{-3}$, $Y_e=0.01$, and $T=10~\mathrm{MeV}$.}
  \label{fig:mc_neut_sat_T}
\end{figure}

Fig.~\ref{fig:mc_neut_sat_T} shows the distribution in the free energy
per baryon at a slightly larger temperature, and the entropy
contribution drops the free energy per baryon in each case. In
comparison to the results from Fig.~\ref{fig:mc_neut_sat}, the LS220
free energy per particle drops more than SFHo or SFHx because its
effective mass (equal to the nucleon mass) is larger than that in
SFHo/x (about 0.7 times the nucleon mass). The effective mass can be
computed from the chiral interaction directly
\cite{holt13prcb,holt16prc}, and close to the Fermi surface it is
found to be nearly equal to the free-space nucleon mass. Second-order
many-body perturbation theory contributions, however, produce a strong
momentum dependence in this region, and averaging around $k=k_F$ the
effective mass is about 0.85 times the free-space nucleon mass. The
distribution in Fig.~\ref{fig:mc_neut_sat_T} is thus larger than LS220
because of the smaller symmetry energy ($S=28.6$~MeV) and larger
effective mass in the LS220 EOS.

\begin{figure}
  \includegraphics[width=0.5\textwidth]{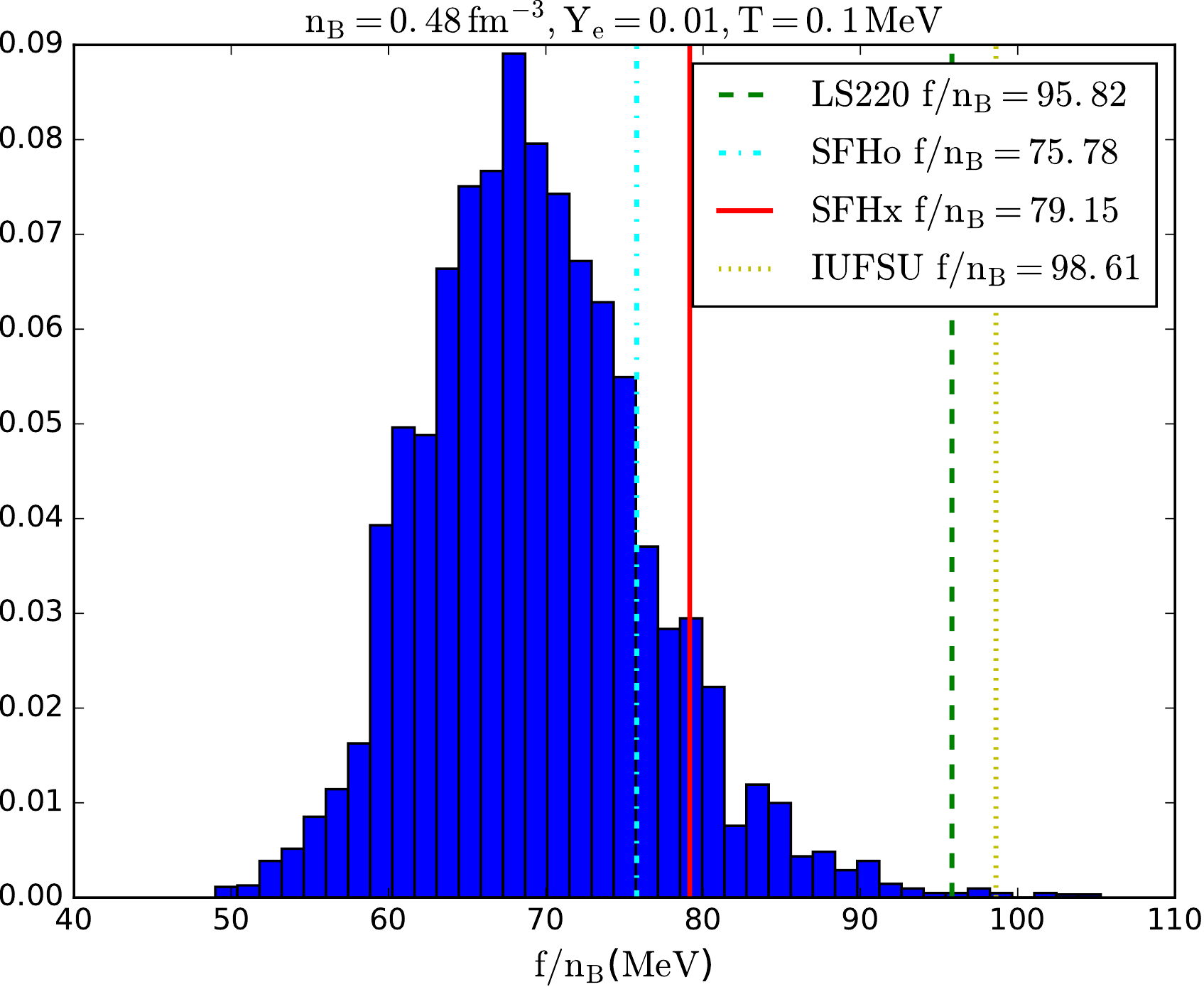}
  \caption{The probability distribution for the free energy per baryon
    at $n_B=0.48~\mathrm{fm}^{-3}$, $Y_e=0.10$, and $T=0.1~\mathrm{MeV}$.}
  \label{fig:mc_hd_neut}
\end{figure}

Fig.~\ref{fig:mc_hd_neut} shows the probability distribution for the
low-temperature neutron-rich matter free energy per baryon
at higher densities. There is clearly a much larger spread in the
free energy per baryon, corresponding to our larger ignorance
regarding the nature of matter at higher density. While all models
LS220, SFHo, SFHx and IUFSU are inside the region suggested by
our parameterization, our distribution leans towards
smaller values of the free energy because of the constraint
from relatively small neutron star radii.

\begin{figure}
  \includegraphics[width=0.5\textwidth]{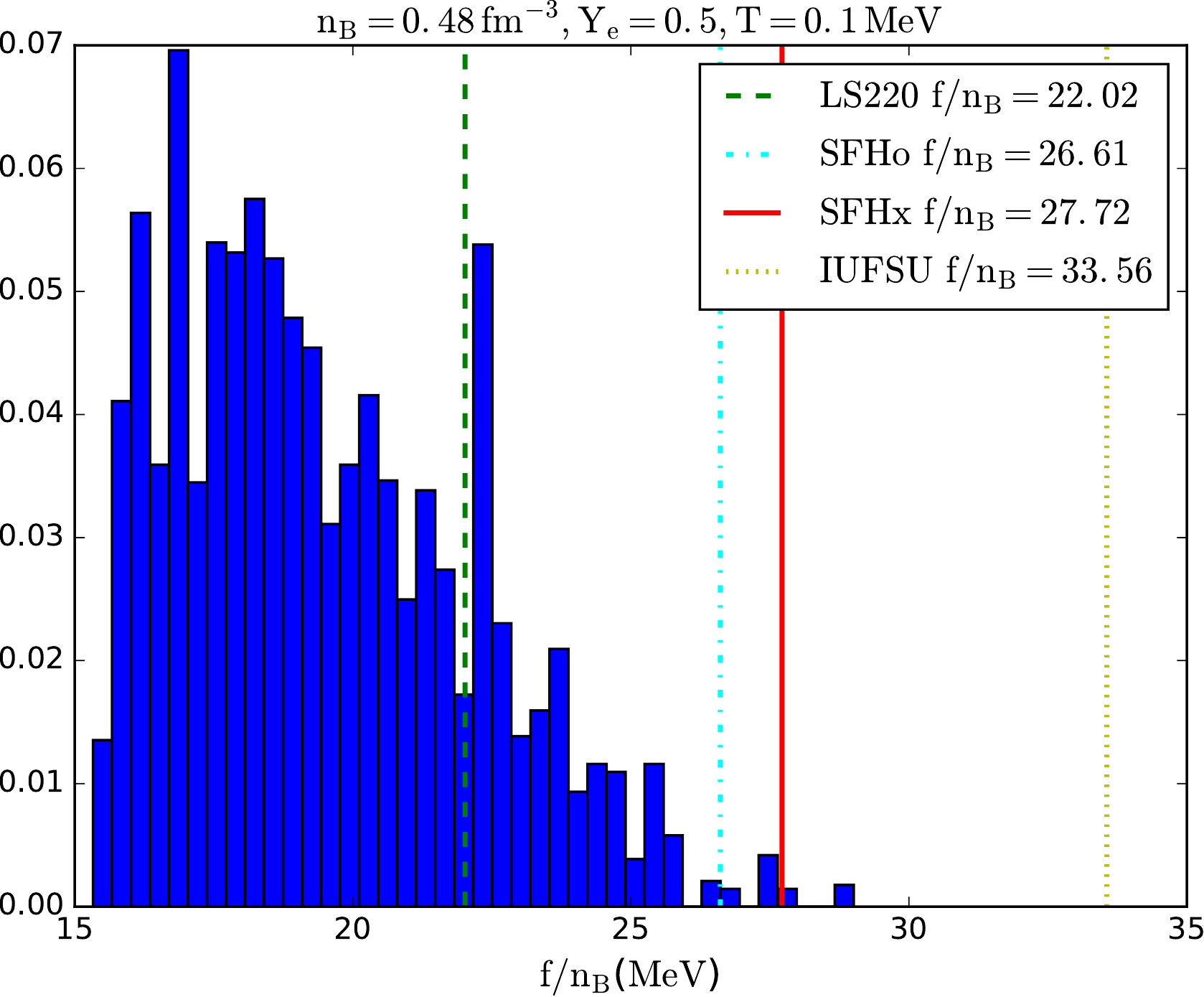}
  \caption{The probability distribution for the free energy per baryon
    at $n_B=0.48~\mathrm{fm}^{-3}$, $Y_e=0.5$, and $T=0.1~\mathrm{MeV}$.}
  \label{fig:mc_hd_nuc}
\end{figure}

Fig.~\ref{fig:mc_hd_nuc} shows low-temperature nuclear matter at
higher densities. The non-smooth nature of the distribution is due to
the small statistics afforded by the limited number of Skyrme models
we have employed. IUFSU suggests a larger free energy here because it
originates in a relativistic mean field model which tends to give
larger pressures than the non-relativistic models like Skyrme. This
region of parameter space is almost entirely unconstrained by
experiment, since it is not possible to make cold isospin symmetric
matter at this density. However, dense isospin-symmetric matter is not
as relevant for this work since simulations are typically neutron-rich
at high density.

\begin{figure}
  \includegraphics[width=0.5\textwidth]{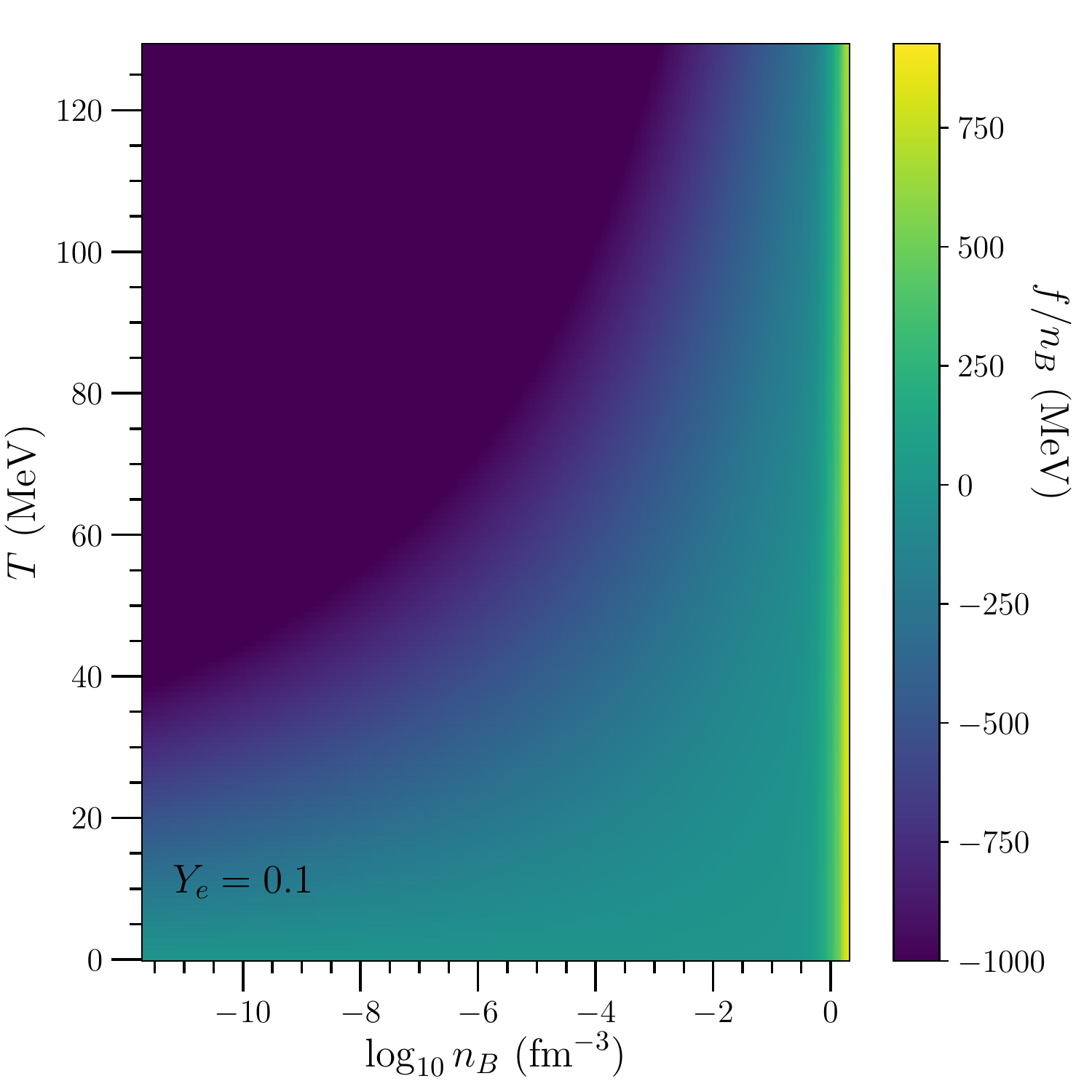}
  \includegraphics[width=0.5\textwidth]{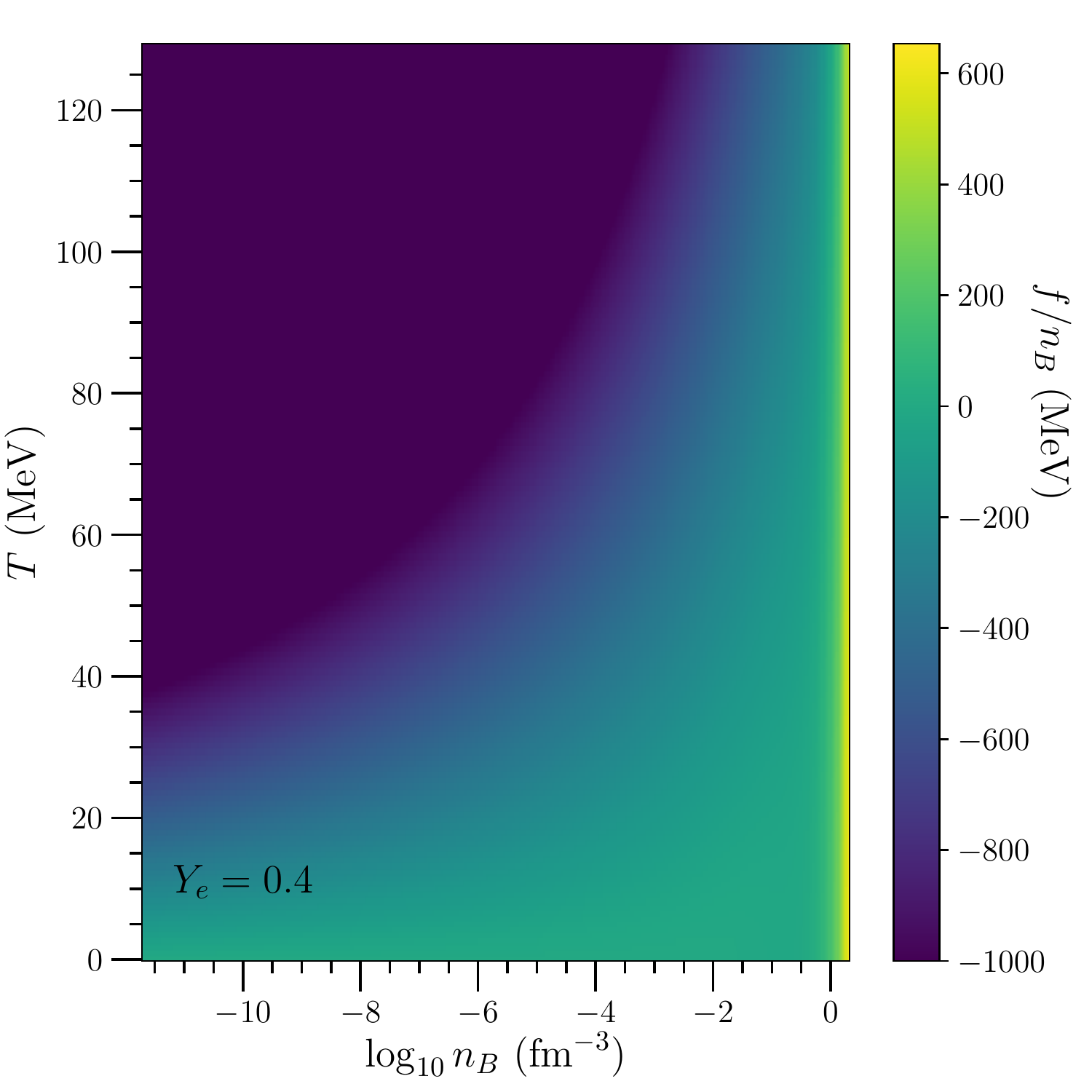}
  \caption{Density plots showing the free energy per baryon for matter
    at $Y_e=0.1$ and $Y_e=0.4$ over the full range of densities and
    temperatures considered in this work for one parameterization.
    Points with a free energy per baryon less than $-$1000 MeV are set
    equal to $-$1000 MeV to make the high-density behavior more clear.
    The main variation in the free energy per baryon from the
    lower-right region to the upper-left region in these plots is
    dominated by the virial contribution to the EOS. The degenerate
    part of the EOS is clear in the large increase in the free energy
    per baryon on the right-hand boundary (at large baryon densities).
  }
  \label{fig:tabYe}
\end{figure}

Fig.~\ref{fig:tabYe} shows the free energy per
baryon for one of our parameterizations as a function of baryon
density and temperature for two electron fractions. Using the
formalism presented in this work, thousands of similar density
plots can be generated with alternate parameterizations. The
largest variation between parameterizations is in the
free energy per particle at high density which changes with
the ``stiffness'' of the EOS.

\begin{figure}
  \includegraphics[width=0.5\textwidth]{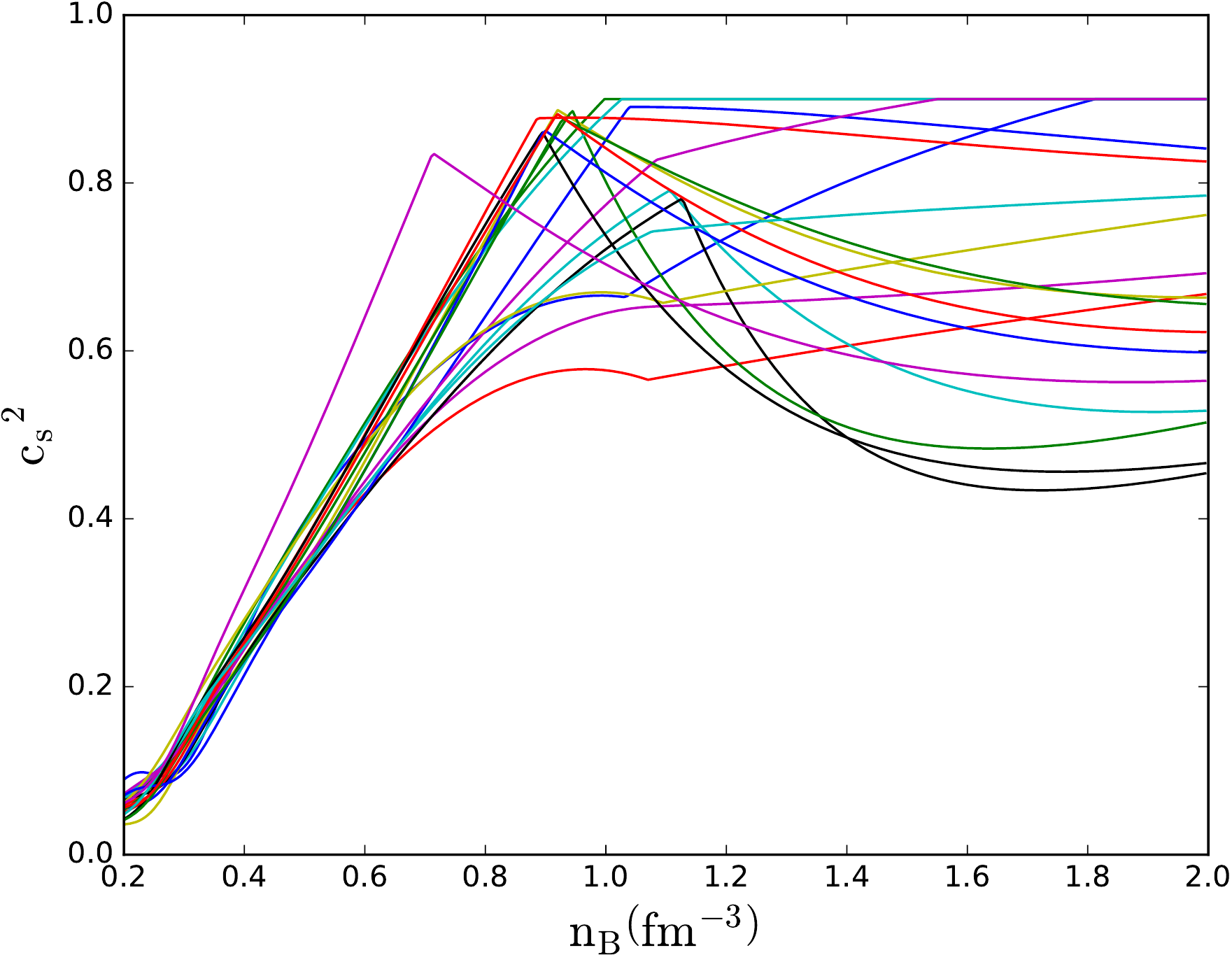}
  \caption{Values of $\mathrm{n_B}$ and $\mathrm{{c_s}^2}$ explored by
    the EOS at $\mathrm{Y_e}=0.1$ and $\tilde{s}$=0.5, The different
    curves represent randomly selected models. At $\mathrm{Y_e}=0.1$,
    most models strongly affected by an decreasing interpolation of
    Neutron star EOS and therefore decrease more dramatically at
    higher density region. }
  \label{fig:cs2a}
\end{figure}

Since accurate neutron star radii have not yet been measured for large
mass neutron stars, the speed of sound of matter at the highest
densities probed in supernova and merger simulations is not
constrained by experiment. We have parameterized this variation with
$\phi$. However, the speed of sound must increase at moderate
densities in order to reproduce the observation of a two solar mass
neutron star. Fig.~\ref{fig:cs2a} shows the behavior of the speed of
sound in neutron-rich matter between $n_B=0.1$ $\mathrm{fm}^{-3}$ and
$n_B=2~\mathrm{fm}^{-3}$ and demonstrates these two regimes. The
speed of sound must increase quickly below $1~\mathrm{fm}^{-3}$ to
ensure that the neutron star maximum mass is sufficiently large, and
the speed of sound at higher densities varies considerably depending
on the value of $\phi$. We restrict the speed of sound to be less than
$c\sqrt{0.9}$ to ensure finite-precision errors in simulations do
not create unphysical sound speeds.

\begin{figure}
  \includegraphics[width=0.5\textwidth]{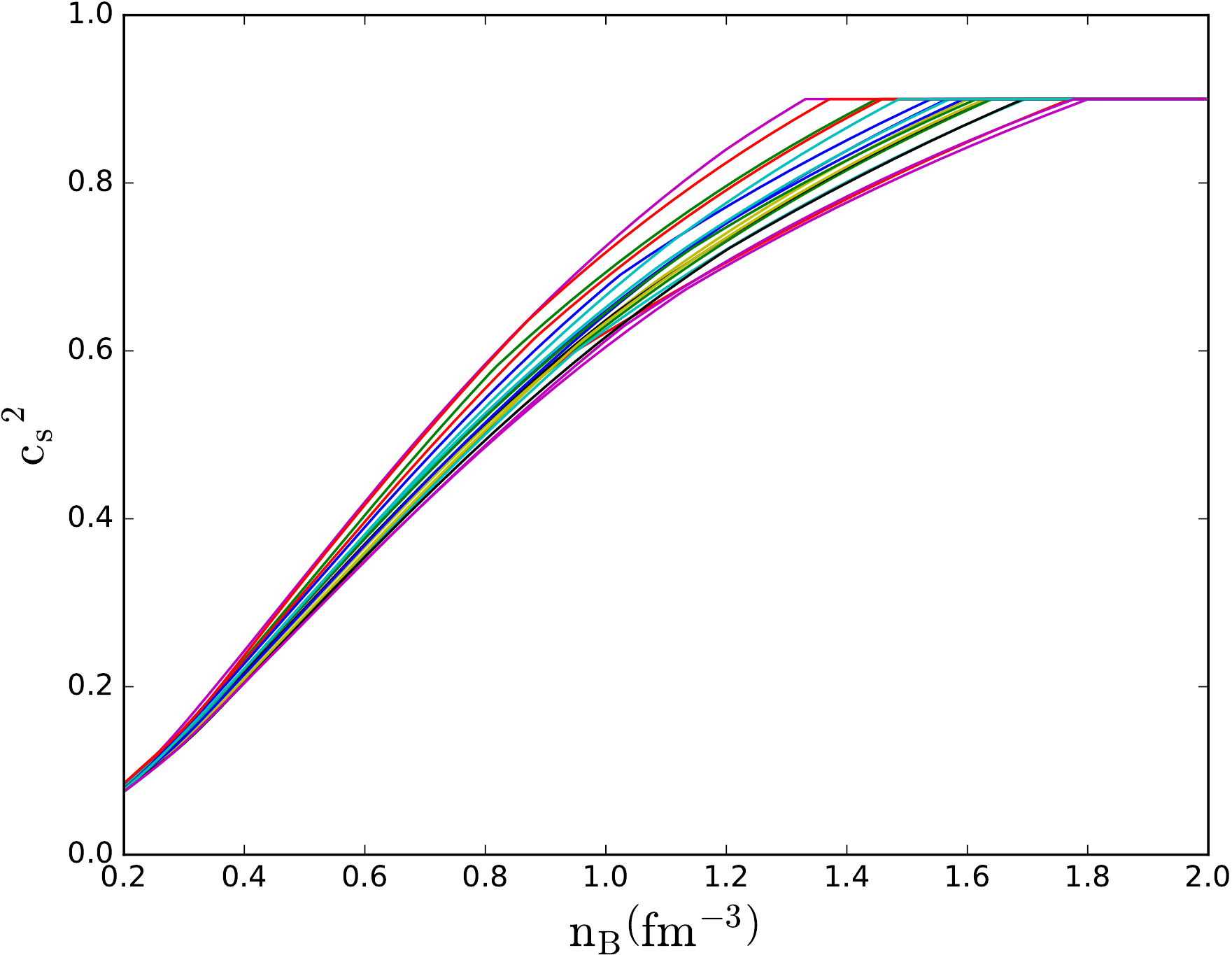}
  \caption{Values of $\mathrm{n_B}$ and $\mathrm{{c_s}^2}$
    explored by the EOS at $\mathrm{Y_e}=0.4$ and
    $\tilde{s}$=0.5, The different curves represent randomly
    selected models. Most of them have two kinks. The first one
    (often very slight) is the residual impact of fixing
    the speed of sound in neutron matter at high densities to $\phi$
    as shown in Fig.~\ref{fig:cs2a}. The second kink
    between $1.3 < n_B^{*} < 1.8~\mathrm{fm}^{-3}$ is due
    to the use of the C\&P prescription to decrease the
    speed of sound above $n_B=n_B^{*}$.}
  \label{fig:cs2b}
\end{figure}

In high-density isospin-symmetric matter, the speed of sound is
dominated by the Skyrme model used for isospin-symmetric matter near
the saturation density. This region is principally constrained to have
a speed of sound less than $c\sqrt{0.9}$ by our implementation of the
prescription from Ref.~\cite{Constantinou17} as described in
section~\ref{s:causal}. This is shown in Fig.~\ref{fig:cs2b}. There is
also a slight residual impact from the modification in the speed of
sound from $\phi$ at values of $Y_e$ which are nearly but not exactly
equal to $1/2$, so there are some slight kinks in the curves in
Fig.~\ref{fig:cs2b} near $n_B=1~\mathrm{fm}^{-3}$.

\begin{figure}
  \includegraphics[width=0.5\textwidth]{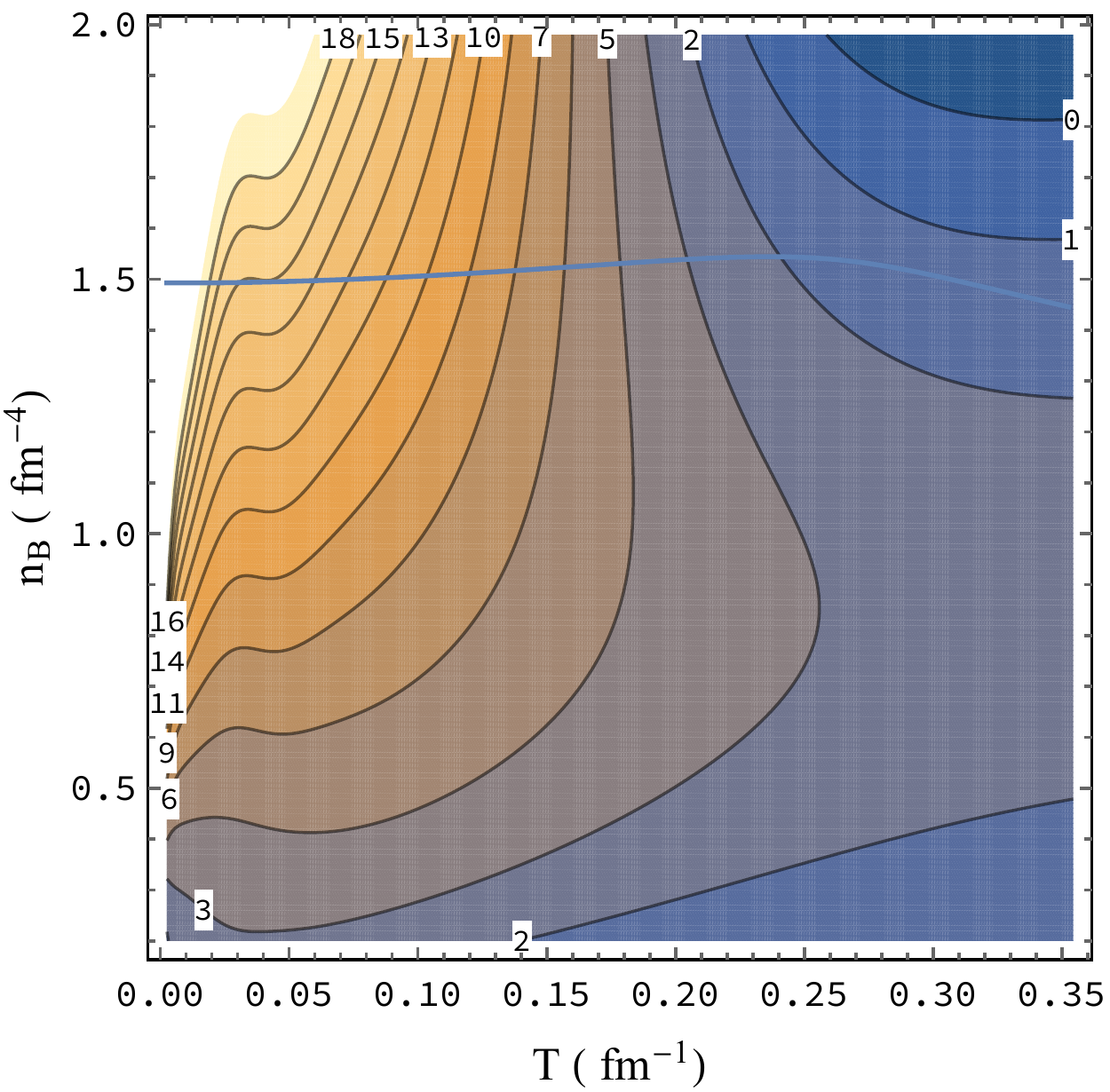}
  \caption{Contour plot of $ds/dT$ as a function of $n_B$ and $T$
    explored by the EOS for one of our EOS parameterizations. The
    speed of sound correction from the C\&P prescription has not yet
    been applied. The blue line indicates the value $n_B=n_B^{*}$.
    The region above this line is replaced with the
    C\&P EOS (including the region where $ds/dT<0$).}
  \label{fig:cs2c}
\end{figure}

At sufficiently high density, the entropy from the chiral EOS begins
decreasing with increasing temperature. Because of the presence of
$ds/dT$ in the speed of sound (see $f_{TT}$ in the denominator of
Eq.~(\ref{eq:ss_final})), this unstable region implies a large speed of
sound. Our use of the C\&P prescription thus cures this instability in
the extrapolated form of the finite-temperature corrections from the
chiral EOS. This is demonstrated in Fig.~\ref{fig:cs2c}, which shows
contours of fixed $ds/dT$. This derivative becomes negative in the
upper right region, but this is always at a density larger than
$n_B^{*}$ where the C\&P EOS takes over.

As a final demonstration that our implementation of the C\&P
prescription generates a continuous EOS, we show the entropy as a
function of density for several fixed temperatures. Our EOS above is
used for densities below $n_B^{*}$ (indicated by the red dot) and the
C\&P EOS is used for densities above $n_B^{*}$. It appears in this
plot that $n_B^{*}$ is independent of temperature, but this is not
exactly true as there is a weak temperature dependence as shown in
Fig.~\ref{fig:cs2c}. We have found that solving Eqs.~(\ref{eq:five}) is
numerically challenging because of the numerical derivatives involved
in computing the speed of sound (we use exact expressions for the
entropy and chemical potentials but numerical differentiation for the
number susceptibilities and other second derivatives of the free
energy). The combination of the numerical derivatives plus the
numerical noise in the Newton-Raphson method used to solve
Eqs.~(\ref{eq:five}) leads to a bit of noise in the entropy at large
densities. Future work will use exact expressions for second
derivatives of the free energy and thus facilitate the correction to
the speed of sound.

\begin{figure}
  \includegraphics[width=0.5\textwidth]{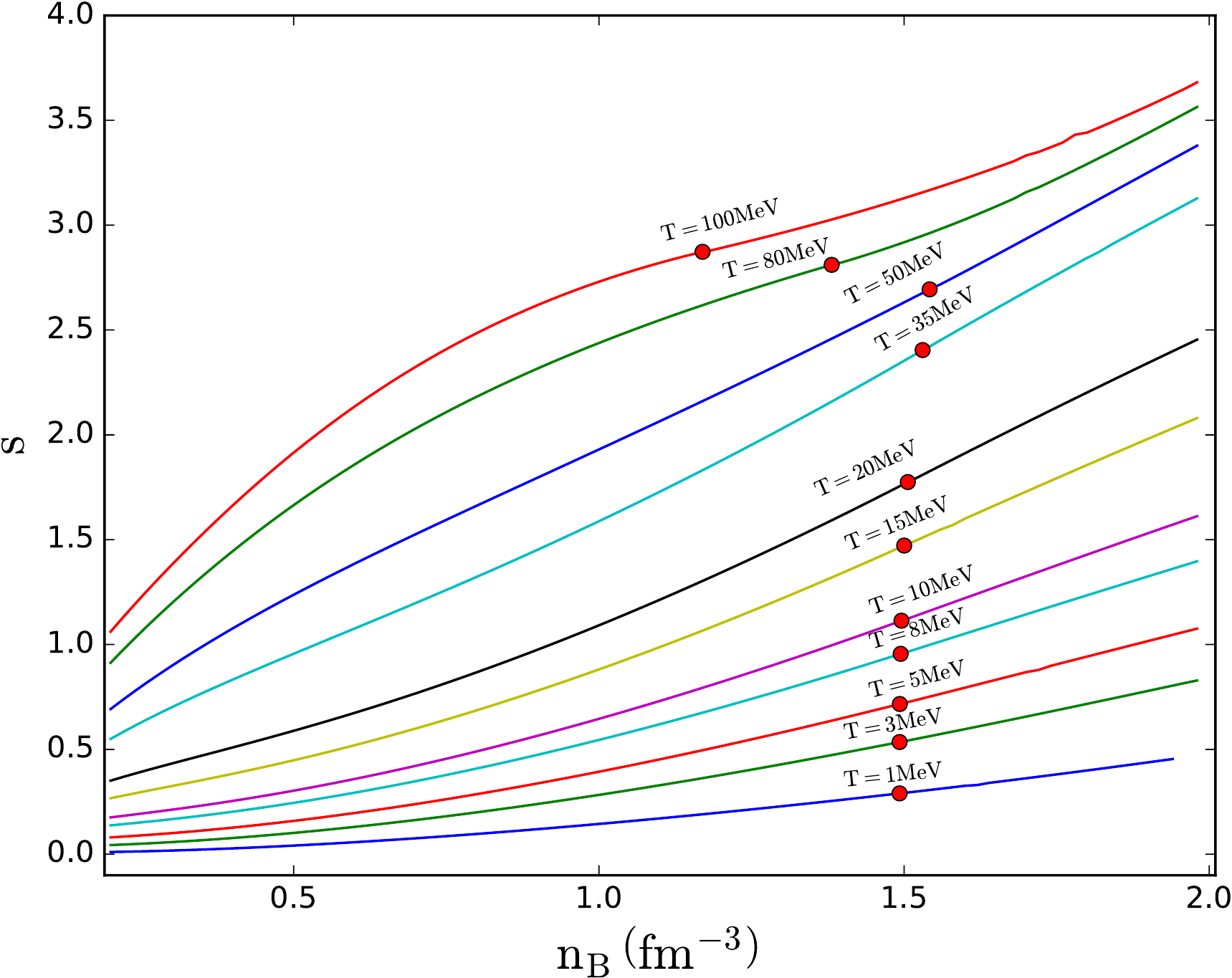}
  \caption{Values of $n_B$ and $s$ explored by the EOS. The dots
    represent boundaries where ${c_s}^2=0.9$ and C\&P solution starts
    to take in effect.}
  \label{fig:dsdT}
\end{figure}

\section{Discussion}

While we have made an attempt to explore the allowed EOS space as much
as possible, there are several regions in which our parameterization
is limited. Variations in the functions $g$ and $h$ could be explored,
but modifications of these functions cannot be too large or they are
likely to create small regions where the EOS has an acausal speed of
sound. A better quantification of the uncertainties in the finite
temperature part of the chiral EOS will be performed in future work.
We have also naively extrapolated the Skyrme model used in
isospin-symmetric matter at zero temperature near the saturation
density to higher densities. There is some experimental constraint on
matter in this region from heavy ion collisions~\cite{Tsang12co}, but
the associated systematic uncertainties are not well-understood. In
any case, dense and isospin-symmetric matter is not often explored in
the neutron-rich matter encountered in neutron stars.

Uncertainties in the EOS may be relevant for the timescale for
core-collapse supernovae to explode after bounce~\cite{Couch13td} and
also, for example, in determining the amount of r-process
nucleosynthesis which occurs in the neutrino-driven
wind~\cite{Roberts12mm}. EOS uncertainties are also relevant for
neutron star mergers, as they dictate the lifetime of hypermassive
neutron star remnants and also the amount of r-process material
ejected~\cite{Sekiguchi15dm}.

Because we use the Markov chain from ``Model A'' in
Ref.~\cite{Steiner15un}, our EOS specifically prefers more moderate
phase transitions, which is appropriate to our assumption that matter
consists only of nucleons and no exotic matter. Thus our
uncertainties at high-density may be underestimated because strong
phase transitions are disfavored. One of the advantages of our
analytical form for the EOS is that our work can be easily generalized
to an EOS which includes exotic matter at high densities if desired.

This article is the first step towards a full quantification of how
microphysical uncertainties may affect core-collapse supernovae and
neutron star mergers. The next step is a full description of nuclei in
the dense matter environment with uncertainties that properly reflect
the relationship between nuclear structure and the underlying
nucleon-nucleon interaction. One way to include nuclei on top of our
EOS for homogeneous matter is to use the framework recently developed
in Ref.~\cite{Schneider17on}. Finally, the EOS uncertainties must be
propagated through to the neutrino opacities. As this uncertainty
quantification matures, the comparison of simulations with data
points, such as GW170817~\cite{Abbott17go} and future nearby
core-collapse supernovae will provide more insight into what models
might be ruled out.

\section*{Acknowledgements}

The authors would like to thank D. Higdon for suppling the samples
from the posterior generated in Ref.~\cite{Kortelainen14}. XD and AWS
were supported by DOE SciDAC grant DE-SC0018232 and NSF grant PHY
1554876, and JWH was supported by NSF grant PHY 1652199. Portions of
this research were conducted with the advanced computing resources
provided by Texas A\&M High Performance Research Computing.

\section*{Appendix I - Speed of Sound for a Multicomponent System}

Using $\varepsilon$ for energy density, $S$ for entropy,
$s$ for entropy density, and $\tilde{s}$
for entropy per baryon, and assuming neutrinos are not trapped,
the speed of sound is (all chemical potentials and energy densities
below include the rest mass contribution even though not explicitly
indicated)
\begin{equation}
c_s^2 = \left( \frac{\partial P}{\partial \varepsilon}
\right)_{\tilde{s},\{ N_i \}}
\, .
\end{equation}
In infinite matter, it is useful to rewrite this derivative in
terms of fixed volume rather than fixed number. 
\begin{equation}
c_s^2 = \left( \frac{\partial P}{\partial \varepsilon} \right)_{S,\{ N_i \}} =
\left( \frac{\partial P}{\partial V} \right)_{S,\{ N_i \}}
\left( \frac{\partial \varepsilon}{\partial V} \right)_{S,\{ N_i \}}^{-1},
\end{equation}
where the second derivative on the right-hand-side of this expression is
\begin{eqnarray}
\left( \frac{\partial \varepsilon}{\partial V} \right)_{S,\{ N_i \}} &=& 
\left[ \frac{\partial  (E/V)}{\partial V} \right]_{S,\{ N_i \}} =
-\frac{1}{V} P - \frac{E}{V^2} 
\nonumber \\
&=& - \frac{P+\varepsilon}{V}
= - \frac{T s + \sum_i \mu_i n_i}{V}
\end{eqnarray}
and the first derivative on the right-hand side is
\begin{eqnarray}
\left( \frac{\partial P}{\partial V} \right)_{S,\{ N_j \}} &=& -
\left( \frac{\partial \varepsilon}{\partial V} \right)_{S,\{ N_j\}} +
S \left[ \frac{\partial (T/V)}{\partial V} \right]_{S,\{ N_j \}} +
\nonumber \\
&&
\sum_i 
N_i \left[ \frac{\partial  (\mu_i/V)}{\partial V} \right]_{S,\{ N_j \}}
\\ &=& -
\left( \frac{\partial \varepsilon}{\partial V} \right)_{S,\{ N_j \}} 
\nonumber \\
&& +S \left[ -\frac{T}{V^2} + \left( \frac{\partial T}{\partial V}
\right)_{S,\{ N_j \}}\right] 
\nonumber \\
&& + \sum_i 
 N_i \left[ -\frac{\mu_i}{V^2} +
\left( \frac{\partial \mu_i}{\partial V} \right)_{S,\{ N_j \}}\right]
\\ &=& \frac{P + \varepsilon}{V} +
S \left[ -\frac{T}{V^2} - \left( \frac{\partial P}{\partial S}
\right)_{\{N_j\},V}\right] 
\\ \nonumber
&& +\sum_i N_i \left[ -\frac{\mu_i}{V^2} -
\left( \frac{\partial P}{\partial N_i}
\right)_{S,\{N_{j\neq i}\},V}\right] \nonumber \\
&=& - S \left( \frac{\partial P}{\partial S}\right)_{\{n_j\},V}
- \sum_i N_i \left( \frac{\partial P}{\partial N_i}
\right)_{S,\{n_{j\neq i}\},V}
\end{eqnarray}
Putting these two results together gives
\begin{eqnarray}
c_s^2 &=& \left[s \left( \frac{\partial P}{\partial s}
\right)_{\{n_j\},V} +
\sum_i n_i \left( \frac{\partial P}
{\partial n_i} \right)_{S,\{n_{j\neq i}\},V}\right]
\nonumber \\
&& \left( T s + \sum_i \mu_i n_i \right)^{-1}.
\end{eqnarray}
To re-express this in terms of derivatives of the free energy,
\begin{eqnarray}
c_s^2 &=& \left\{s \left[ \frac{\partial (\sum_i \mu_i n_i - f)}
{\partial s} \right]_{\{n_j\},V} \right.
\\ \nonumber
&& \hspace{-.3in}\left. +\sum_i 
n_i\left[ \frac{\partial  ( \sum_k \mu_k n_k - f)}{\partial n_i}
\right]_{s,\{n_{j\neq i}\},V}\right\} \left(
T s + \sum_i \mu_i n_i \right)^{-1}.
\end{eqnarray}
For the sum over $k$,
all densities are constant except for $n_i$, thus 
\begin{eqnarray}
&& \sum_i 
n_i \frac{\partial}{\partial n_i}
\left( \sum_k \mu_k n_k - f \right)_{s,\{n_{j\neq i}\},V}
\\ \nonumber
&=& \sum_i n_i \frac{\partial}{\partial n_i}
\left( \sum_{k\neq i} \mu_k n_k + \mu_i n_i -f
\right)_{s,\{n_{j\neq i}\},V} \nonumber \\
&=& 
\sum_i \left[ \sum_k n_k \left(\frac{\partial \mu_k }
{\partial n_i}\right)_{s,\{n_{j\neq i}\},V} + \mu_i \right.
\\ \nonumber
&& \left. -\left(\frac{\partial f}{\partial n_i}\right)_{s,\{n_{j\neq i}\},V}
\right].
\end{eqnarray}

To compute this we need
\begin{eqnarray}
\left(\frac{\partial f}{\partial n_i}\right)_{s,\{n_{j\neq i}\},V} &=&
\left(\frac{\partial f}{\partial n_i}\right)_{\{n_{j\neq i}\},T,V} 
\\ \nonumber
&& +\left(\frac{\partial f}{\partial T}\right)_{n_B,\{n_{j\neq i}\},V}
\left(\frac{\partial T}{\partial n_i}\right)_{\{n_{j\neq i}\},s,V}
\\ \nonumber
&=& \mu_i - s \left(\frac{\partial T}{\partial n_i}\right)_{\{n_{j\neq i}\},s,V}
\nonumber \\
\left(\frac{\partial \mu_k}{\partial n_i}\right)_{s,\{n_{j\neq i}\},V} &=&
\left(\frac{\partial \mu_k}{\partial n_i}\right)_{\{n_{j\neq i}\},T,V} 
\\ \nonumber
&& +\left(\frac{\partial \mu_k}{\partial T}\right)_{n_i,\{n_{j\neq i}\},V}
\left(\frac{\partial T}{\partial n_i}\right)_{\{n_{j\neq i}\},s,V} 
\\ \nonumber
&& = f_{n_i n_k} + f_{n_k T}
\left(\frac{\partial T}{\partial n_i}\right)_{\{n_{j\neq i}\},s,V}
\end{eqnarray}
which requires
\begin{eqnarray}
\left(\frac{\partial T}{\partial n_i}\right)_{\{n_{j\neq i}\},s,V}
&=& -\left(\frac{\partial s}{\partial n_i}\right)_{\{n_{j\neq i}\},T,V}
\left(\frac{\partial s}{\partial T}\right)_{\{n\},V}^{-1}
\\ \nonumber
&=& -f_{n_i T}/f_{TT}
\end{eqnarray}
Finally, we get
\begin{eqnarray}
c_s^2 &=& \left\{
- \left(\frac{s}{f_{TT}}\right) \left( \sum_i n_i f_{n_i T}+s \right) \right.
\\ \nonumber
&& \left. + \sum_i n_i \left[ \sum_k n_k \left(f_{n_i n_k}- f_{n_k T}
f_{n_i T} f_{TT}^{-1}\right) 
- s f_{n_i T} f_{TT}^{-1}\right]
\right\} 
\\ \nonumber
&& \left( T s + \sum_i \mu_i n_i \right)^{-1} \nonumber \\
&=& \left[
\sum_i \sum_k n_i n_k \left(f_{n_i n_k}- f_{n_k T}
f_{n_i T} f_{TT}^{-1}\right) \right.
\\ \nonumber
&& \left .- 2\sum_i s n_i f_{n_i T} f_{TT}^{-1}
- s^2 f_{TT}^{-1} \right] \left(
T s + \sum_i \mu_i n_i \right)^{-1}.
\label{eq:ss_final}
\end{eqnarray}

\bibliographystyle{apsrev}
\bibliography{paper} 

\end{document}